\documentclass[aps,prd,showpacs,nofootinbib,floats,floatfix,preprintnumbers,groupedaddress,twocolumn]{revtex4}
\usepackage{bm}
\usepackage{latexsym}
\usepackage{dcolumn}
\usepackage{amsmath,amsfonts,amssymb}
\usepackage{graphicx,epsfig}
\usepackage{amsmath}
\usepackage{hyperref}
\usepackage{graphicx,epstopdf}
\hypersetup{
	colorlinks   = true, %Colours links instead of ugly boxes
	urlcolor     = blue, %Colour for external hyperlinks
	linkcolor    = blue, %Colour of internal links
	citecolor   = red %Colour of citations
}

{\scriptsize }

\begin{document}
\title{Diffeomorphism symmetries near a  timelike surface in black hole spacetime}
\author{Mousumi Maitra\footnote {\color{blue} maitra.91@iitg.ac.in}}
\author{Debaprasad Maity\footnote {\color{blue} debu@iitg.ac.in}}
\author{Bibhas Ranjan Majhi\footnote {\color{blue} bibhas.majhi@iitg.ac.in}}

\affiliation{Department of Physics, Indian Institute of Technology Guwahati, Guwahati 781039, Assam, India
}

\date{\today}
%%%%%%%%%%%%%%%%%%%%%%%%%%%%%%%%%%%%%%%%%%

\begin{abstract}
Recently symmetries of gravity and gauge fields in the asymptotic regions of spacetime have been shown to play vital role in their low energy scattering phenomena. Further, for the black hole spacetime, near horizon symmetry has been observed to play possible role in understanding the underlying degrees of freedom for thermodynamic behaviour of horizon. Following the similar idea, in this paper, we analyzed the symmetry and associated algebra near a timelike surface which is situated at any arbitrary radial position and is embedded in black hole spacetime. In this paper we considered both Schwarzschild and Kerr black hole spacetimes. The families of hypersurfaces with constant radial coordinate (outside the horizon) in these spacetimes is timelike in nature and divide the space into two distinct regions. The symmetry algebra turned out to be reminiscent of Bondi-Metzner-Sach (BMS) symmetry, found in the asymptotic null infinity.  
\end{abstract}

\maketitle
%%%%%%%%%%%%%%%%%%%%%%%%%%%%%%%%%%%%%%%%%%%%%%%%%%%%%%%%%%%%%%%%%%
\section{Introduction}

Symmetries play fundamental role in the development of the fundamental laws of nature. Thanks to the celebrated Noether's theorem, which says that for every continuous global symmetry of a physical system, there exits associated conserved charges. These global symmetries are typically represented by the finite-dimensional group of transformation and associated Lie algebra. In the context of gravitational theory, it is the diffeomorphism symmetry which plays fundamental role. The associated Noether charges \cite{Wald:1993nt,Iyer:1994ys} are believed to be encoded into the thermodynamics properties of gravity \cite{Bekenstein:1973ur,Bekenstein:1974ax,Bekenstein:1972tm}. One of the most striking object of gravity is black hole which is manifestly endowed with thermodynamics properties specially attributed to the horizon \cite{Wald:1993nt,Iyer:1994ys,Bekenstein:1972tm,Bekenstein:1973ur,Bekenstein:1974ax,Bardeen:1973gs,Hawking:1974sw,Hawking:1976de,Gibbons:1977mu,Hawking:1982dh}. Horizon is a null hyper-surface located at a special point in radial coordinate of the black hole spacetime. Out of infinite dimensional diffeomorphism symmetry, it has been shown to exist a special class called asymptotic symmetry which forms a closed Virasoro algebra \cite{Strominger:1997eq,Carlip:1999cy,Carlip:1998wz,Majhi:2011ws,Majhi:2012nq,Majhi:2012tf,Majhi:2013lba,Majhi:2014lka,Majhi:2015tpa,Majhi:2017fua,Bhattacharya:2018epn} (an extensive list of works in this direction can be obtained from reference list of \cite{Majhi:2011ws}). An important thermodynamic quantity namely entropy of the horizon can be expressed in terms of the central charge of this algebra by using well known Cardy formula \cite{Cardy:1986ie}. This approach was first formulated in the seminal work by Brown and Henneaux for the (1+2) dimensional gravity theory \cite{Brown:1986nw}. Later it was proved that the entropy and various other thermodynamic quantities can be attributed to any generic null surface \cite{Parattu:2013gwa,Chakraborty:2015aja,Bhattacharya:2018epn,Dey:2020tkj}.

In recent years the understanding of asymptotic symmetries has received widespread interests in the realm of infrared properties of gravity, particularly focusing on scattering phenomena near the null boundaries of spacetime \cite{Weinberg:1965nx}-\cite{Campiglia:2015qka} (for a review, see \cite{Strominger:2017zoo}). It has been shown that these low energy scattering phenomena can be beautifully described by aforementioned non-trivial group of infinite-dimensional symmetries first appeared in the pioneering works by Bondi, van der Berg, Metzner and Sachs (BMS) \cite{Bondi:1962px}-\cite{Barnich:2013sxa}. In these works, the attempt was to figure out the isometries of asymptotically flat spacetime at null infinity. To their surprise the symmetry group turned out to be an infinite dimensional one with Poincar\'e as sub group. Apart from usual Poincar\'e transformation, the additional infinite number of generators of symmetry transformation are coined as ``supertranslations''. The crucial point in all these discussion was related to the boundary conditions. Significant efforts have been devoted in specifying the appropriate set of boundary conditions on the gravitational field configuration. One of the popular procedure is to identify the fall off behavior of the background metric components  expressed in Bondi-Sachs coordinates at null infinity. The basic criterion of identifying those boundary conditions was that under arbitrary diffeomorphism the form of the background metric must be preserved at its asymptotic null boundaries. Subsequently  analogous asymptotic symmetry analysis has been  explored near another null surface namely, horizon of black hole \cite{Donnay:2015abr}-\cite{Eling:2016xlx}. Further analysis have been done near both extremal and non-extremal null surfaces in a more general setting which include Lanczos-Lovelock (LL) gravity, non-gravitational fields such as U(1) electromagnetism \cite{Maitra:2018saa}. Also, the symmetry parameters are shown to behave as Goldstone modes which leads to thermal nature of the horizon at the semi-classical level \cite{Maitra:2019eix}.

As has been pointed out already most of the studies on asymptotic symmetry has been performed either at asymptotic null infinity or near the black hole horizon. A natural question arises, {\em what will be the physical implication of the diffeomorphism symmetries on a physical boundary located in the bulk spacetime?} To explore the full power of this asymptotic symmetries in understanding the nature of gravity, we believe it is timely to  
explore the asymptotic symmetries near the families of non-null hypersurface positioned at any arbitrary radial coordinate in the bulk spacetime manifold. The hypersurface can be thought of as domain wall dividing the spacetime into two regions. In our present discussion we will consider a timelike hypersurface situated at a finite radial distance outside the black hole horizon. 

To motivate further with regard to our choice of the timelike boundary in the bulk spacetime, it is important to mention the recent developments on the correspondence between gravity and fluid dynamics, where timelike surface away from the horizon play important part. The interesting connection has been established between Einstein field equations of gravity and Navier-Stokes (NS) equation of any fluid  on a timelike surface \cite{Price:1986yy}-\cite{Dey:2020ogs}. 
Wherein the $(p+2)$-dimensional metric solution of Einstein's equations of motion describes a $(p+1)$-dimensional non-relativistic incompressible fluid on the timelike surface. This is obtained from the conservation Brown-York surface stress energy tensor on the timelike hypersurface. 
Interesting connection has been uncovered between the fluid dynamics on the timelike membrane and infinite set of symmetries and associated conserved charges which are similar to the BMS symmetries and charges \cite{Penna:2015gza}. All these important results motivate us to perform the detail investigation on the diffeomorphism symmetries of the timelike surface under consideration.

%%%%%%%%%%%%%%%%%%%%%%%%%%%%%%%%%%%%%%%%
%%%%%%%%%%%%%%%%%%%%%%%%%%%%%%%%%%%%%%%
A timelike hypersurface defined by the radial coordinate $r=r_c > r_H$ is assumed to be situated outside the black hole horizon. In the present paper we consider both Schwarzschild and Kerr black hole spacetimes. First step would be to express the metric near the surface in Gaussian normal coordinate. In order to do that we have followed the methodology outlined in \cite{Booth:2012xm}. 
Once the metric is derived we will follow the usual procedure for the asymptotic symmetry analysis. By imposing appropriate boundary conditions near the surface we  determine the diffeomorphism symmetry parameters. The boundary conditions are such that the form of the metric near the timelike surface must remain invariant under arbitrary diffeomorphism. The conserved charges due to the diffeomorphism symmetries are calculated in a spacelike subspace of this timelike surface and the associated symmetry algebra are constructed thereafter.

\section{The Gaussian normal coordinate system: a brief review}

In this section we give a quick overview on the construction  of the Gaussian normal coordinate (GNC) system which will be used in the later part of our analysis. These coordinate system is  suitable to explain physical quantities on an $(p+1)$-dimensional hypersuface $\mathcal{S}$, embedded in the $(p+2)$-dimensional manifold $\mathcal{N}$. For the present discussion we consider sub-manifold $\mathcal{S}$ as a timelike hypersurface. One can choose a unique vector $N^a$ in the tangent space of the manifold $\mathcal{N}$ in such a way that the vector will be orthogonal to all the vectors in the tangent space of the submanifold $\mathcal{S}$. So for $\mathcal{S}$, we can normalize $N^a$ such that $N^a  N_a =1$; i.e. $N^a$ is spacelike corresponding to signature $(-,+,+,+)$ of the metric. Thus the vector $N^a$ is unit normal to $\mathcal{S}$. 

Now  we can construct unique geodesic with tangent $N^a$ emanating from an arbitrary point $\mathcal{P}$ on the aforesaid hypersurface. So the geodesic must pass through another arbitrary point $\mathcal{Q}$ situated in the small neighbourhood of the hypersurface $\mathcal{S}$. 
In this setup one defines a coordinate system in the form $( x^1,x^2,x^3, \rho)$, among them ($ x^1,x^2, x^3)$ denote the coordinates of the point $\mathcal{P}$ on the surface. On the other hand fourth coordinate $\rho$ labels the point $\mathcal{Q}$ which lies on the geodesic, very near to $\mathcal{S}$. Here $\rho$ is a spacelike coordinate while any one of $(x^1,x^2,x^3)$ is timelike and rest are spacelike coordinates. So the GNC system $( x^1,x^2,x^3, \rho)$ will be well defined in the small neighbourhood of the point $\mathcal{P}$ in such a way that the congruence of spacelike geodesics will be orthogonal to the hypersurface. The spacetime metric in the neighbourhood of the surface can be expressed in the following form (more details in this regard can be found in \cite{MORALES}):
\begin{eqnarray}
ds^2 = g_{ab}dx^a dx^b = {d\rho}^2 + h_{\mu \nu} dx^{\mu} dx^{\nu}~.
\label{GN}
\label{M1}
\end{eqnarray}
Here $\mu$, $\nu$ indices are defined on the hypersurface $\mathcal{S}$ whose induced metric is given by  $h_{\mu \nu}$. Whereas, $a,b$ correspond to spacetime indices. Here we have metric coefficient $g_{\rho\rho}=1$ due to the normalization of the vector field $N^a$. Furthermore $g_{\rho \mu}=0$ since vector $N^a$ is orthogonal to the surface.  Hence in this newly formed coordinate system, $\rho=$ constant hypersurface describes timelike hypersurface $\mathcal{S}$ in GNC. 

Next in the upcoming sections we will express Schwarzschild and Kerr geometry in this coordinate system in order to study the symmetry properties of the timelike surface which is situated at any arbitrary radial position in these black hole spacetime background.

\section{Schwarzschild black hole}

\subsection{Schwarzschild in GNC} 
The Schwarzschild metric in original Schwarzschild coordinates is given by,
\begin{eqnarray}
&&ds^2 = -(1-2M/r) dt^2 + \frac{dr^2}{(1-2M/r)} +  r^2 d\Omega^2 ~, 
\label{swarz}
\end{eqnarray}
where, $d\Omega^2 = (d\theta^2 + \sin^2\theta d\phi^2)$. Since our main objective is to study the symmetry of this spacetime near a timelike surface, we shall express the above metric in GNC near this hypersurface. We choose $\mathcal{S}$ as $r=r_c$ (constant) hypersurface lies outside the horizon $r=2M$. Below we shall follow the procedure, described in the last section, to express metric (\ref{swarz}) in GNC (more detail in this regard can be found in \cite{Booth:2012xm}).

 The unit spacelike normal vector to $\mathcal{S}$ for Eq. (\ref{swarz}) is given by, $N^a = (0, \sqrt{1-\frac{2M}{r}},0,0)$. The spacelike geodesic congruence  that cross the surface orthogonally will posses $N^a$ as a tangential vector near the surface. Near the surface we can parametrize the geodesics with affine parameter $\rho =r-r_c$.   The timelike surface is identified as $\rho =0$. Therefore, the  geodesics curve $X^a (\rho)$ with $X^a = (t,r,\theta,\phi)$ and orthogonal to $S$ at the intersection, can be expressed through the following Taylor series expansion,
 \begin{eqnarray}
X^a (\rho) \approx   X^a \vline_{\rho= 0} +\rho \frac{dX^a}{d\rho} \vline_{\rho=0} + \frac{{\rho}^2}{2} \frac{d^2 X^a}{d\rho^2} \vline_{\rho= 0} +\dots~.
\label{geodesic}
\end{eqnarray} 
The first term in the right hand side is identified as $X^a \vline_{\rho=0} =( t,r_c,\theta, \phi) $. The second one is given by the tangent vector to the curve, $\frac{dX^a}{d\rho} \vline_{\rho=0} = N^a\vline_{\rho=0}$.
The third term can be evaluated by considering the geodesic equation $N^{a} \nabla_{a} N^{b}=0$ at $\rho=0$. This yields
\begin{eqnarray}
\frac{d^2 r}{d \rho^2} \vline_{\rho= 0} =-\Gamma^{r}_{b c} N^{b} N^{c} \vline_{\rho= 0}=\frac{M}{r_c^2}~.
\label{geodesic1}
\end{eqnarray}
Where all the other component equations identically vanish.
Using these in Eq. (\ref{geodesic}) and keeping upto second order in $\rho$, we find the transformation of coordinates from $(t,r,\theta,\phi)$ to $(t',\rho, \theta',\phi')$ as,
\begin{eqnarray}
&& t=t';\,\,\,\ r = r_c + \rho \sqrt{1-\frac{2M}{r_c}}+ \rho^2 \frac{ M}{2 r_c^2}~;
\\
\nonumber
&& \theta= \theta';\,\,\,\  \phi =\phi'~.
\label{tranf}
\end{eqnarray}
Application of these transformations into Eq. (\ref{swarz}) leads to the following form of the metric:
\begin{eqnarray}
ds^2 = g_{\rho \rho}(\rho) d \rho^2 + g_{tt}(\rho) dt^2 + g_{AB}(\rho,x) dx^A dx^B~,
\label{metric}
\end{eqnarray}
where the argument $x$ corresponds to the angular coordinates ($\theta,\phi$). Now the metric coefficients can be obtained by simply using tensor transformation and keeping up to $\mathcal{O}(\rho)$ as follows
\begin{eqnarray}
g_{ \rho \rho}(\rho) &=& g_{r r} \frac{\partial r }{ \partial \rho}\frac{\partial r }{ \partial \rho},  \nonumber
\\
= \Big(\sqrt{1-\frac{2 M}{r_c}} &+& \rho \frac{ M}{r_c^2} \Big)^2 \frac{1}{\Big(1- \frac{2M}{(r_c + \rho \sqrt{1-2M/r_c}+ \rho^2 \frac{ M}{r_c^2})}  \Big)}\nonumber
\\ \label{app} 
&\simeq& 1+\mathcal{O}(\rho^2),
\\
g_{tt}(\rho) &\simeq& -\Big(1-\frac{2 M}{r_c}\Big) - \Big(\frac{2 M\sqrt{1-\frac{2 M}{r_c}}}{r^2_c}\Big) \rho~,
\nonumber
\\
g_{\theta\theta}(\rho, x^A) &\simeq& \Big(r^2_c +  2 r_c \rho \sqrt{1-\frac{2 M}{r_c}}\Big),
\nonumber
\\
g_{\phi\phi}(\rho, x^A) &\simeq& \Big(r^2_c +  2 r_c \rho \sqrt{1-\frac{2 M}{r_c}}\Big)\sin^2 \theta.
\label{comp}
\end{eqnarray}

Note that in the sufficiently small neighbourhood of the surface the metric Eq. (\ref{metric}) assumes the form Eq. (\ref{GN}) to the leading order in $\rho$. Near the timelike hypersurface $r=r_c$, the asymptotic form of the metric will be exactly the same as given in (\ref{metric}) expressed in Gaussian normal coordinates. Our primary goal would be study the properties of near $r=r_c$ symmetries analogous to the near horizon or near null infinity. Hence we will impose appropriate asymptotic conditions on the variation of the metric coefficients of equation (\ref{metric}).
  %%%%%%%%%%%%%%%%%%%%%%%%%%%%%%%%%%%%%%%%%%%%%%%%%%%%%%%%%%%%%%%%%%%%%%%%%%%
\subsection{Symmetries near the surface}
The key ideas in identifying the asymptotic symmetries are intertwined with the appropriate  set of boundary conditions on the variation of the metric coefficients. Due to diffeomorphism the structure of the metric (\ref{metric}) is assumed to remain same in our region of interest. Generically the asymptotic conditions on the metric coefficients are of two different categories. In one category, the gauge fixed conditions are strictly assumed to be invariant under diffeomorphism transformation $x^{a} \rightarrow x^{a} + \zeta^{a}$ and those are  
\begin{equation}
\pounds_\zeta  g_{\rho \rho}= 0, \ \pounds_\zeta  g_{\rho t}= 0, \ \pounds_\zeta  g_{\rho A}= 0 ~.\label{con1}
\end{equation}
In the above expression $\pounds_\zeta$ implies the Lie derivative along the vector field $\zeta^a$. We call the above set as ``strong'' conditions. These four conditions Eq. (\ref{con1}) essentially provide residual diffeomorhism transformation generators. The functional form of those generators can possibly be restricted by the other category of transformations under which the metric coefficients are assumed to change with the appropriate fall-off conditions as 
 \begin{equation} 
\pounds_\zeta  g_{tt} \approx \mathcal{O}(1); \ \pounds_\zeta  g_{tA} \approx \mathcal{O}(\rho^2),\pounds_\zeta  g_{AB} \approx \mathcal{O}(1).
\label{con2}
\end{equation}
Analogous to the analysis near the null boundary, the asymptotic conditions on the metric variation are given by four gauge conditions Eq. (\ref{con1}) and remaining  fall-off conditions Eq. (\ref{con2}) defined near the timelike boundary in the present discussion. The weak fall off condition for $g_{tt}$ is postulated considering the leading order behaviour of the metric coefficient. The off-diagonal  component $g_{tA}$ is absent in the original Schwarzschild solution , hence we have assumed the fall-off rate of $g_{tA}$ as $\mathcal{O}(\rho^2)$. Whereas the fall-off condition of $g_{AB}$ is imposed as  $\mathcal{O}(1)$ depending on  the leading order structure of the metric. In the next part of the analysis we will show that for the obtained form of the solution of the vector field $\zeta^a$, the Lie variations of  $g_{tA}$ and $g_{AB}$ automatically vanishes. 
We now first solve the gauge fixed condition Eq. (\ref{con1}) to identify the appropriate symmetry parameter and the form of the solutions are as follows,
\begin{eqnarray}
&& \zeta^{\rho} = T(t, x)~;
\nonumber
\\
&& \zeta^t = 1/(1-2 M/r_c) \int \partial_t T(t, x^A) d\rho + F (t, x)~;
\nonumber
\\
&& \zeta^A =  -g^{AC} \int \partial_C T \ d\rho +R^A(t,x)~;
\label{sol}
\end{eqnarray}
where $T ,F$ and $R^A$ are the integration constants. These are the residual diffeomorphism symmetry parameters under which the gauge choice remains the same.
To proceed further, we chose $R^A =0$ as Schwarzschild spacetime does not posses any angular momentum to begin with. Further, we also set  $T(t, x) = 0$ which keeps the position of the hypersurface intact at $\rho=0$. Therefore, we are left with one diffeomorpism parameter $F$ which is similar to the `super-translation'  discussed in the literature for the asymptotic null boundary and null horizon. Now one can show that the Lie variation of the component $g_{tA}$ will give only one term, $g_{tt} \partial_A F$ which remains non-zero near the surface $S$ at $\rho=0$. Therefore the functional form of $F$ needs to be further constrained by the following condition so that the structure of the metric near $S$ does not changes,
\begin{eqnarray}
\partial_A F =0~\implies F = F(t). \label{constraint1}
\end{eqnarray}

Finally we can write symmetry parameters very near to the surface as;
\begin{equation}
\zeta^{\rho} = 0 = \zeta^A ~; \,\,\,\
\zeta^t =  F(t)~.
\label{solF}
\end{equation}
Hence for the above condition (\ref{constraint1}) on parameter $F$ and for the solution set (\ref{solF}) of  $\zeta^a$ the Lie variation of the  metric component $g_{tA}$ and $g_{AB}$ will be zero automatically. So in (\ref{con2}) there will be fall-off condition on the component $g_{tt}$ only.  We have left with the diffeomorphism along time direction. This feature is completely different from what was obtained for the case of a general null-surface. (see e.g. \cite{Maitra:2018saa}). In the later part of the discussion we will discuss that the derived form of  $\zeta^a$ is such that all the metric coefficients preserve their asymptotic form, while the black hole parameter mass gets modified with a time dependent term. This phenomena is similar to the near horizon symmetries or asymptotic symmetries near infinity where it has been observed that under infinitesemial diffeomorphism structure of the metric near the boundaries does not change, but the black hole parameters, like mass, can be modified.
Analogous to the near horizon symmetry, we call  $F(t)$  as `supertranslation' parameter. {\em Unlike near horizon symmetries \cite{Donnay:2015abr} and asymptotic null infinity BMS symmetries \cite{Sachs:1962zza}, symmetry generator near the timelike hypersurface embedded in the Schwarzschild spacetime, the  `supertranslation' parameter is not arbitrary function of spacial coordinates ($\theta,\phi$), rather it turned out to be dependent only on time.}

\subsection{Algebra of symmetry parameter}
Now we want to construct symmetry algebra followed by the symmetry generator constructed in the last section.  We have only one non-zero component of $\zeta^a=(F(t),0,0)$ which will generate the symmetry algebra of the residual symmetry  transformation. Since the $g_{tt}$ component of the metric does not diverge near $\rho =0$, the periodicity of Euclidean time near the surface in principle is infinite. Therefore, associated Fourier mode of $F$ will be  continuous in frequency space ,
\begin{eqnarray}
&& F(t) = \int^{\infty}_{-\infty} d \omega ~ \alpha(\omega) e^{-i \omega t}
\nonumber\\ 
&=& \int^{\infty}_{-\infty} d \omega ~ \alpha(\omega) \bar{F}(\omega,t)~. 
\label{Fm}
\end{eqnarray}
Here $\alpha(\omega)$  represents individual Fourier mode of $ F(t)$. The kernel function $\bar{F}(\omega,t) = \exp{(-i\omega t)}$  will form the complete orthogonal basis with continuous frequency $\omega$.

Now Let us consider two parameters $ \zeta^a_1= (F_1,0,0)$ and $ \zeta^a_2= (F_2,0,0)$ associated with the symmetry generators. 
The standard Lie bracket algebra between these two diffeomorphism vectors is given by \cite{Barnich:2010eb},
\begin{equation}
[\zeta_1, \zeta_2]_{M} = [\zeta_1, \zeta_2] -  \delta^g_{\zeta_1} \zeta_2 - \delta^g_{\zeta_2}  \zeta_1~,
\label{modified}
\end{equation}
where,
\begin{equation}
[\zeta_1, \zeta_2]^x= \zeta_1^a \partial_a \zeta_2^x -\zeta_2^a \partial_a \zeta_1^x~.
\label{Lie}
\end{equation}
Here $\delta^g_{\zeta_1} \zeta_2$ denotes the variation in $\zeta_2$ under the variation of the metric induced by $\zeta_1$. Second and third terms in the definition of Lie bracket (\ref{modified}) are considered in order to take into account the variation of the vector fields due to the higher order variation of the metric. But we will ignore higher order effect on the variation of diffeomorphism vectors. Therefore, Eq. (\ref{Lie}) will be our main equation to construct the symmetry algebra. 
In terms of $F$, the Lie bracket takes the following form,
\begin{eqnarray}
[\zeta_1,\zeta_2]^t = [F_1,F_2]= F_1 \partial_t F_2-F_2 \partial_t F_1~. \label{bracket}
\end{eqnarray}
In terms of Fourier modes, the left hand side becomes,
\begin{equation}
[F_1,F_2]= \int^{\infty}_{-\infty} \int^{\infty}_{-\infty}  d\omega_1 d \omega_2  {\alpha} (\omega_1,\omega_2) [\bar{F}(\omega_1,t),\bar{F}(\omega_2,t)]~,\label{left}
\end{equation} 
with ${\alpha} (\omega_1,\omega_2)=\alpha(\omega_1)\alpha(\omega_2)$, and the right hand side of the equation Eq. (\ref{bracket}) becomes,
\begin{eqnarray}
&&F_1 \partial_t F_2-F_2 \partial_t F_1 
\nonumber
\\ 
&=& \int^{\infty}_{-\infty} \int^{\infty}_{-\infty}  d\omega_1 d \omega_2  \alpha (\omega_1) \alpha(\omega_2) [-i(\omega_2 -\omega_1)]
\nonumber
\\
 &\times&\bar{F}(\omega_1+ \omega_2,t)~.
\label{right}
\end{eqnarray}
Comparing both the equations Eq. (\ref{left}) and Eq. (\ref{right}), the algebra among the basis functions becomes,
\begin{eqnarray}
i [\bar{F}(\omega_1,t),\bar{F}(\omega_2,t)]= (\omega_2 -\omega_1) \bar{F} (\omega_1+ \omega_2,t).\nonumber\\ \label{braketM}
\end{eqnarray}
Here the important point to mention is that although the diffeomorphism symmetry parameter defined on the timelike surface in Schwarzschild spacetime came out to be dependent only on time, but the aforementioned bracket algebra (\ref{braketM}) is similar to the that obtained for a generic null surface (\cite{Maitra:2018saa}).
%\end{equation}
 %%%%%%%%%%%%%%%%%%%%%%%%%%%%%%%%%%%%%%%%%%%%%%%%%%%%%%%%%%%%%%%

\subsection{ charges on the surface}
For any diffeomorphsim invariant gravity theory conserved charges corresponding to symmetries play crucial role in understanding the thermodynamics of the black hole. Innumerable studies have been done where some specific set of diffeomorphism symmetric charges have been shown to be related to the black hole entropy \cite{Iyer:1994ys}-\cite{Majhi:2015tpa}. In all these works, the charges have been calculated on the two dimensional cross-section (given by $t=$ constant spacelike subspace) of the horizon three dimensional $r=$ constant null hypersurface. Here we are dealing with timelike surface which is three dimensional. In the present context we shall calculate the charge at a particular instant of coordinate time. 

 In Einstein gravity conserved charges due to diffeomorphism symmetry on a two-surface (either on horizon or at infinity) is expressed in the following way \cite{Paddy},
\begin{equation}
Q[\zeta]=\frac{1}{2}\int d\Sigma^{ab}J_{ab}~,
\label{Q}
\end{equation}
where $J_{ab}$ is the Noether potential, given by
\begin{eqnarray}
J_{ab} = \frac{1}{16\pi G} (\nabla_a \zeta_b- \nabla_b \zeta_a)~.
\label{Jab}
\end{eqnarray}
The integration is being done on the two-surface. 
Before proceeding towards computing this quantity associated with the symmetries near our time like surface discussed above, let us make few comments on the meaning of $Q$ when assigned on the cross-section of this surface. It is well known that the Noether current $J^a$ associated with the diffeomorphism symmetry $x^a\rightarrow x^a+\zeta^a$ for Einstein-Hilbert action satisfies  covariant conservation equation for arbitrary $\zeta^a$ (for instance, see \cite{Paddy} and Appendix B of \cite{Majhi:2014lka}). Here $\zeta^a$, in principle, is completely arbitrary. Hence the conserved charge can naturally be $Q_t = \int_{\mathcal{V}} d\Sigma_a J^a$, where $d\Sigma_a$ is the volume element on a $t=$ constant surface. Now $J^a$ can be expressed as $J^a = \nabla_b J^{ab}$ where $J^{ab}$ is anti-symmetric. Using Stokes' theorem charge $Q_t$ is transformed into following closed surface integral $Q_t=\oint_{\partial\mathcal{V}} d\Sigma_{ab}J^{ab}$, where $d\Sigma_{ab}$ is a surface element of the closed boundary $\partial\mathcal{V}$ which encloses the volume $\mathcal{V}$.
Depending upon the system under study boundary may contains multiple disconnected closed surfaces. For instance black hole spacetime has two natural set of boundaries at the two dimensional cross-section of the null infinity and of the horizon. If the boundary surfaces are disconnected, one can compute the charge for every individual surface and understand their properties. However, one should keep in mind that individual boundary charge may not be conserved in general unless one takes into account all the surface contributions.

In the present discussion  we have considered a $t=constant$ slice of $r=r_c$ timelike hyper surface which act as a two dimensional boundary between horizon and infinity. Here $Q_t$ can be written as $Q_t = \Big[Q_t(r=r_H) + Q_t(r=r_c,-N^a)\Big] + \Big[Q_t(r=r_c,N^a) + Q_t(r\rightarrow\infty)\Big]$, where the expression in the first part corresponds to region within horizon and $r_c$ while the second part is connected to region between $r_c$ to $\infty$. It is expected that $Q_t(r=r_c,-N^a)$ and $Q_t(r=r_c,N^a)$ must cancel each other, because the normal on $r=r_c$ surface are same but in opposite direction with respect to two regions. However, the value of individual terms may be non-vanishing. This suggests that in principle one can define a unique quantity $Q\sim \int d\Sigma_{ab}J^{ab}$ on any $t=r=constant$ two dimensional cross-section which can be thought of part of the conserved charge $Q_t$ if the observer is confined in either one side of the surface. In this sense $Q$ alone should not be a conserved quantity. However, we use the usual nomenclature as ``charge" here in a very weak sense as this part is not alone conserved.  This charge can be interpreted as the induced charge on an arbitrary time like hypersurface associated with a special class of diffeomorphism under study.

Now natural question arises, what symmetry does this charge should correspond to?
For any arbitrary diffeomorphism, as we stated earlier, the charge Eq. (\ref{Q}) can be defined in principal on any arbitrary surface. This is defined for any arbitrary diffeomorphism. Therefore the choice of $\zeta^a$ (the gauge choice) from a particular condition will construct a charge which is conserved in the sense of our previous discussion. Even if one finds this diffeomorphism from a local (gauge) symmetry of metric, the total corresponding charge which is sum of contributions from different parts of the boundary, still conserved as this is by construction conserved for any arbitrary diffeomorphism. In fact given a specific form of the diffeomorphism associated with the underlying symmetry the formula yields  the well known charges near the horizon or null-infinity.  This procedure has been shown to be yielded horizon entropy in the context of Virasoro algebra (e.g. see \cite{Carlip:1999cy,Carlip:1998wz,Majhi:2011ws,Majhi:2012tf}). Note that even if the conserved charge here is due to gauge symmetry, its value on a part of the boundary may be non-vanishing. However, in this paper, what would be the physical characteristics and how to compute the charge on the arbitrary time like surface are the questions we have asked and explored in detail.  
%		 definition of `conserved quantities' can be formulated with the help of asymptotic killing vector fields arising from boundary preserving symmetries. Here the boundary can be the null infinity or black hole horizon where  diffeomorphism of the vector fields does not specify exact symmetries, rather they are large gauge transformation of spacetime. In the present context we have considered a timelike hypersurface which is not essentially a physical boundary of the spacetime. 
		 Hence the charges are calculated for a class of diffeomorphism symmetries which which preserves the form of the metric near the timelike surface under study. Once $Q$ is defined as above, on any two dimensional cross-section (like, hypersurface defined by both $t=$ and $r=$ are constants),
one can compute the associated diffeomorphism symmetry transformation which keeps the metric structure near the surface under study unchanged, following the usual procedure. One usually  computes  $\zeta^a$ from a physically motivated boundary conditions on metric coefficients $g_{ab}$ as just mentioned above. Using this subset of diffeomorphisms, constraint by certain physical boundary conditions on $g_{ab}$, one usually calculates $Q$. This well known procedure has been used  in the case of horizon as well as asymptotic infinity structure preserving symmetries. In both cases it has been observed that structure of the metric remains unchanged, but the black hole parameters, such as mass, angular momentum can be modified.

In the present paper we are calculating this $Q$ on the cross-section of a timelike surface $r=r_c$, denoted by $t=constant$ slice following the same procedure. We also observed that under suitable redefinition,  the black hole parameters changes (e.g. in the case of Schwarzschild, mass has been changed to $M\rightarrow M + (M-\frac{r_c}{2})(\partial_t F)$). However this change may  be thought of as diffeomorphic change of the black hole mass.

In this context, as a side remark, it may be mentioned that since our interest is on the $t=constant$ slice of timelike hypersurface and $Q$ can be associated on any two dimensional surface, it is not mandatory to exists a horizon in the spacetime, as least as far as the definition of $Q$ is concerned. The only restriction has to be that the metric should be solution of diffeomorphism invariant gravity theory, like GR and in that sense the same can be applied for a star spacetime solution.

In the present situation, as mentioned above, we shall calculate $Q$ on the one side of  $\rho=0$ surface at a particular time slice. Therefore this subspace is two dimensional and can be indicated by two unit normals: one is timelike and another is spacelike. Hence this particular subspace is in general spacelike, but not the two dimensional spacelike slice of the horizon  as $r_c\neq 2M$. The surface element on the two-surface is
\begin{eqnarray}
d\Sigma^{ab} =d^2 x \sqrt{\sigma} (M^a n^b-M^b n^a)~,
\label{surface}
\end{eqnarray} 
where $\sigma$ is the determinant of the induced metric on the hypersurface under consideration. $M^a$ is the unit spacelike normal vector directed outward and defined on the timelike surface ($\rho=constant$), and is given by, $M^a=(0,1,0,0)$. Similarly $n^a$ is the unit timelike normal on the spacelike hypersurface ($t=constant$) taking the following form,
\begin{eqnarray}
n^t= \frac{1}{\sqrt{1-2 M/r_c}} - \rho \Big( \frac{M}{(r_c^2-2 Mr_c)}\Big )~.
\label{N}
\end{eqnarray}
%\begin{eqnarray}
%&& M^{\rho} = \frac{\sqrt{(1-2 M/(r_c + \rho \sqrt{1-2 M/r_c}))}}{(\sqrt{(1- 2 M/r_c)} +\rho M/r_c^2)}\label{M}
%N^t= -1/ \sqrt{(1-2 M/(r_c + \rho \sqrt{1-2M/r_c})}
%\end{eqnarray}
 
 %At first non-vanishing components of the vectors are calculated in original coordinates. Then these are expressed in newly defined coordinates (\ref{tranf}) with the help of tensor transformation rules,
%\begin{eqnarray}
%X'^a =\frac{\partial x'^a}{\partial x^b} X^b
%\end{eqnarray}
%These computation is based on the transformation of coordinates (\ref{tranf})starting from the original metric (\ref{swarz}).

As the spacelike surface is situated at $\rho=0$ the only non-vanishing component of the surface element in Eq. (\ref{surface}) is, 
\begin{eqnarray} \label{sigmatro}
d\Sigma^{t \rho} =-d^2x \sqrt{\sigma} n^t M^{\rho}.
\end{eqnarray}
 Using Eq. (\ref{solF}) and Eq. (\ref{sigmatro}) one obtains $Q$ to be,
\begin{eqnarray}
Q[F] =   \frac{M F(t)}{2 G}~.
\label{QF}
\end{eqnarray}
As expected form the spherically symmetric background, the charge $Q$ comes out to be independent of the position of the timelike surface under study. For Kerr black hole background, we will see this does not hold true. Note that it depends on a particular $t=constant$ slice of our preferred hypersurface $r=r_c$. In this case $Q$ can have leakage on this spacelike slice if one moves with time. But it may be noted that similar leakage in $Q$ can be found on the same $t=constant$ slice of the actual physical boundaries (at $r\rightarrow\infty$ and horizon) and also in other side of $r_c$ surface. It may happen that all these leakages are such that the collective quantity vanishes. This feature is not new as it happens for charges on the horizon for general diffeomophism vector $\zeta^a$ (for example, see Eq. $(13)$ of \cite{Majhi:2012tf}; also see \cite{Majhi:2017fua}). The reason is obvious. As mentioned earlier, the covariantly conserved Noether current $J^a$ and consequently the anti-symmetric potential $J^{ab}$ are defined for any arbitrary $\zeta^a$, even it can be time dependent. Therefore the computation of charge on a particular boundary $\partial\mathcal{V}$ can be in general time dependent. However, if one includes all part of $\partial\mathcal{V}$, total charges must be conserved by construction.
 
%which is the well known expression of Bekeinstein-Hawking entropy of the black hole horizon \cite{Bekenstein}. 
%The symmetry analysis on a non-null surface has gained importance  because of its usefulness in null horizon surface symmetry prescription.  A well known approach for handling the symmetry quantities on null surface is to consider it as the limit of series of non-null surfaces. Then the physical quantities are derived on this newly formed surface, at the end a limit is taken by which the physical quantities on that surface moves to that of a null surface. Our present analysis is Based on this prescription. Nevertheless what we found in (\ref{QF}) may be more interesting. The conserved charges come out to be independent of the position of the surface under study.

Next in order to compute bracket algebra between charges we express Eq. (\ref{QF}) in terms of the Fourier transform of the `supertranslation' parameter given in (\ref{Fm}):
\begin{eqnarray}
Q[F]= \frac{ M}{2 G } \int^{\infty}_{-\infty} d \omega ~ \alpha(\omega) \bar{F}(\omega,t)~.
\label{QF1}
\end{eqnarray}
The Fourier mode of the charge is defined as 
\begin{eqnarray}
Q[\bar{F}(\omega,t)]= \frac{ M}{2 G} \bar{F}(\omega,t)~.
\label{Qbar}
\end{eqnarray}
%and,
%%\begin{equation}
%Q[F_2]= \int^{\infty}_{-\infty} d \omega_n \ \alpha(\omega_n) Q[F_n(\omega_n,t)]
%\end{equation}
%Now substituting (\ref{Qbar}) in (\ref{QF1}) we get the following expression,
%\begin{eqnarray}
%Q[F]= \frac{M}{2G} \int^{\infty}_{-\infty} d \omega \ \alpha(\omega) Q[\bar{F}(\omega,t)]\label{QF2}
%\end{eqnarray}
Therefore, the algebra among the charges directly follows from the that of the `supertranslation' $F$ as follows,
\begin{eqnarray}
[Q[\zeta_1],Q[\zeta_2]]= \pounds_{\zeta_1} Q[\zeta_2]~,
\label{ch}
\end{eqnarray}
which leads to,
\begin{eqnarray}
[Q[F_1], Q[F_2]]= \pounds_{F_1} Q[F_2]=  \frac{M}{2 G} [F_1,F_2]
\end{eqnarray}
Using Eq. (\ref{braketM}), the final 
%\begin{eqnarray}
%i \frac{M}{ G} [\bar{F}(\omega_1,t),\bar{F}(\omega_2,t)]= (\omega_2 -\omega_1) \frac{M}{G} \bar{F} (\omega_1 + \omega_2,t).\nonumber\\ \label{ch1}
%\end{eqnarray}
expression becomes,
\begin{eqnarray}
i [Q[\bar{F}(\omega_1,t)], Q[\bar{F}(\omega_2,t)]] &=& (\omega_2 - \omega_1)\nonumber\\
&\times& Q[\bar{F}(\omega_1 +\omega_2,t)].
\end{eqnarray}
%NOw in order to have symmetry analysis 

Before we end the discussion, let us try to investigate whether  the thermodynamic interpretation of the `super-translation' charge Eq. (\ref{Qbar}) can be found in a more general setting.    
%One can show that for a particular value of the function $F(t)$, the above expression (\ref{QF}) becomes proportional to the area of the timelike surface. If we choose $F(t)=  2 \pi r^2_c / M$, then one finds from (\ref{QF}) as
% \begin{equation}
 %Q[F]= Q_{r_c}= \frac{A_{r_c}}{4G}~,
 %\end{equation}
 %where $A_{r_c}=4\pi r_c^2$ is the area of our subspace of timelike hypersurface. For this $\alpha(\omega)$ can be derived by the inverse Fourier transform  of (\ref{Fm}). This is given by
 %\begin{eqnarray}
 %\alpha(\omega)=\frac{1}{2\pi}\int_{-\infty}^{+\infty}dt F(t)e^{i\omega t} = \frac{2 \pi r_c^2}{M}\delta(\omega)~,
 %\label{A}
% \end{eqnarray}
 %and correspondingly $\bar{F}=1$. \textcolor{red}{need to understand this}
Let us start with the general static, spherically symmetric metric of the form
 \begin{equation}
 ds^2=-f_1(r)dt^2 + f_2(r) dr^2 +r^2(d\theta^2+\sin^2\theta d\phi^2)~.
 \label{B1}~
 \end{equation}
 In this case the metric near timelike hypersurface $r_c$ in GNC is determined to be as
 \begin{eqnarray}
 ds^2 &=& -[f_1(r_c) + f_1'(r_c)\sqrt{1/f_2(r_c)}~\rho]~dt^2 + d\rho^2
 \nonumber
 \\ 
 &+& [r_c^2+2r_c\sqrt{1/f_2(r_c)}~\rho]~d\Omega^2~.
 \label{B2}
 \end{eqnarray}
 The above can be obtained by using the following transformations,
 \begin{eqnarray}
 &&r=r_c+\rho\sqrt{1/f_2(r_c)}+ \frac{\rho^2}{4 f_2^2} f_2'(r_c); 
 \nonumber
 \\
 && t\rightarrow t; \,\,\,\ \theta\rightarrow \theta; \,\,\,\ \phi\rightarrow\phi~.
 \end{eqnarray}
Now like earlier the charge for the symmetries near $r_c$, given by (\ref{solF}), for $F(t)=1$ at a particular time turns out to be
\begin{equation}
Q = \frac{A_c}{16\pi G} \frac{f_1'(r_c)}{\sqrt{f_1(r_c) f_2(r_c)}} ~,
\label{B3}
\end{equation}
where $A_{r_c}=4\pi r_c^2$ is the area of our subspace of timelike hypersurface.

We shall now show that the above result may be interpreted as the amount of heat content in the hypersurface under study at a particular instant of time. This can be straight forwardly seen for the simple Schwarzschild case, where the metric coefficient is $f_1 = 1/f_2 = 1-(2M/r)$. The local gravitational acceleration  at the radial coordinate $r_c$ is given by $g(r_c) = M/r_c^2 =f'(r_c)/2$. Hence, the choice of global symmetry generator, $F(t) =1$, yields
\begin{equation}
Q=\frac{A_c}{4G}\frac{g(r_c)}{2\pi}~.
\label{B4}
\end{equation}
For general spherically symmetric black hole Eq. (\ref{B1}), it can be proved that local acceleration takes the from $g(r_c) = f'(r_c)/{\sqrt{4f_1(r_c) f_2(r_c)}}$. A discussion regarding this has been presented in Appendix \ref{App1}. Hence, the quantity $Q$  may be interpreted as the {\it local} heat content on the hypersurface at a fixed time if one  identifies the surface entropy $S_c \propto A_c $ and local temperature $T_c \propto g(r_c)$. The temperature is identified as the Unruh temperature which is directly connected with the local gravitational acceleration $g(r_c)$. 
%and according to equivalence principle locally a uniformly accelerated frame with this acceleration can mimic gravity. Then
Hence, with respect to the local observer hovering near the $r=r_c$ hypersurface will see the surface as thermal object with the aforementioned Unruh temperature $T_c$.

The same conclusion can also be obtained though an interesting perspective considering the Tolman relation \cite{Tolman:1930zza}-\cite{Tolman}, which gives rise to temperature gradient of a thermodynamic equilibrium system in the gravitational field. The relation entails $T(r) \sqrt{ g_{ab} k^a k^b} = T(r) \sqrt{-f_1(r)}= T_0$, where $k^a =(1,0,0,0)$ is the timelike Killing vector of the background under consideration and $T(r)$ is called Tolman temperature. $T_0$ is arbitrary constant equilibrium temperature. Using the relation $f_1' = -2f_1(T'/T)$ and $F(t)=1$ which precisely corresponds to the time-like Killing vector $k^a$, we obtain the following interesting relation,                              
\begin{equation}
Q= -\frac{1}{2\pi}\frac{A_c}{4G} \left(\frac {T_{0}}{T}\right)\frac{\hat{n}^a\nabla_a T}{T}\Big|_{r_c}~
\label{B55}
\end{equation}
where we have chosen the normal to $r=$ constant surface in such a way that $\hat{n}_a=(0,\sqrt{f_2},0,0)$ and consequently $\hat{n}^a = (0,1/\sqrt{f_2},0,0)$.
 Now it has been shown in \cite{Santiago:2018kds} that a classical gas of radiation under Newtonian acceleration satisfies $\nabla_aT/T = -g_a$. Therefore the temperature gradient part of (\ref{B55}) leads to gravitational acceleration along the normal to hypersurface and hence we obtain (\ref{B4}).  
It would be interesting to explore this relation Eq. (\ref{B55}) further in the context thermodynamic origin of gravitational theory.

 Hence for a globally defined diffeomorphism vector $F(t) =1$, the charge induced on the timelike hyper-surface can be expressed as Eq. 	(\ref{B55}), for which we are able to provide a distinct thermodynamic interpretation. However, for arbitrary time dependent globally defined function of $F(t)$, we are unable to give any physical interpretation. However, one should remember the fact that near the horizon or near null infinity, the charges corresponding to those diffeomorphism are still not well understood. 
 % The idea is analogous to the near horizon symmetry analysis where the conserved charge corresponding to zero modes of the supertranslation parameter (which is a constant) is shown to be related with the thermodynamics quantities of black hole. Whereas for different choices of the symmetry parameter (for all other modes) charges can be explained as the supertranslation or superrotation hair of the black hole.} 

\subsection{Comparison the results in Minkowski background}

So far we have tried to explore the symmetry of the non-null hypersurface situated at $r_c > r_H$ in the Schwarzschild black hole background. In this section we consider its flat space limit $M\rightarrow 0$. One expected outcome naturally follows from the Eq. (\ref{QF1}) is that the expression of charge $Q$ vanishes on the timelike hypersurface. This can also be straightforwardly calculated considering a specific set of the timelike surface at any arbitrary radial position in the Minkowski background. In spherical polar coordinates the metric is 
\begin{eqnarray}
&&ds^2 = - dt^2 + dr^2 + r^2 (d\theta^2 + \sin\theta^2 d\phi^2)~. \label{flat}
\end{eqnarray}
Following the same procedure as discussed in detail we can obtain the induced metric on a  time like surface defined by $r=r_c$ in Gaussian normal coordinates as,
\begin{eqnarray}
&& ds^2 =  d \rho^2 - dt^2 + (r^2_c +  2 r_c \rho )(d\theta^2 +\sin^2 \theta d\phi^2).\nonumber\\ \label{flatMetric}
\end{eqnarray}
%\textcolor{blue}
The Lie variation of the gauge conditions and other metric components as fall-off conditions will remain the same as given in Eq. (\ref{con1}) and Eq. (\ref{con2}). Therefore, the solution of the gauge choices will be same as found in Eq. (\ref{sol}) with the limit $ M \rightarrow 0$. The constraints coming from the fall-off condition will be same as before Eq. (\ref{constraint1}). Hence symmetry algebra will remain same Eq. (\ref{braketM}) as for the Schwarzschild black hole background. However, as mentioned in the beginning, our explicit computation and also obvious from Eq. (\ref{QF1}) the associated  charge $Q$ becomes zero.
As we have already pointed out in the previous section, the super-translation charge is directly related to an important thermodynamics quantity called heat content associated with the global parameter $F=1$. Hence, our explicit computation for the flat spacetime gives consistent null results for the same.

%%%%%%%%%%%%%%%%%%%%%%%%%%%%%%%%%%%%%%%%%%%%%%%%%%%%%%%%%%%%%%%%%%%%%
  
\section{Kerr background}
\subsection{Kerr in GNC}
We will now extend earlier discussion for a timelike surface located outside the horizon of Kerr black hole. The Kerr metric in Boyer-Lindquist coordinate is expressed as
\begin{eqnarray}
&&ds^2 = -\frac{\Delta -a^2 \sin^2 \theta}{\Sigma} dt^2 
\nonumber
\\
&-& \frac{4 a M r \sin^2\theta}{\Sigma} dt d\phi + \frac{\Sigma}{\Delta} dr^2 +\Sigma d\theta^2 
\nonumber
\\
&+& \frac{\sin^2\theta [(r^2 +a^2)^2 -a^2 \sin^2\theta \Delta]}{\Sigma}d\phi^2~, 
\label{kmetric}
\end{eqnarray}
where, $\Delta= r^2+a^2-2Mr $ and $\Sigma= r^2 +a^2 \cos^2\theta$~.

 %We define
We will follow the same methodology as elaborately discussed for the Schwarzschild black hole. Our task is to express the metric in the neighbourhood of the time like surface $r=r_c > r_H$ in Gaussian normal coordinate in the form of Eq. (\ref{M1}). The detail procedure in this regard has been discussed in the last section in case of Schwarzschild spacetime.
 
 In the Kerr background spacetime, the non-vanishing component of the unit space like normal vector $N^a$ to the hypersurface under consideration  is given by,
\begin{equation}
 N^r = \sqrt{ \frac{(r^2+a^2-2Mr )}{{(r^2 +a^2 \cos^2\theta)}}}~.
 \end{equation}
Now following the same procedure discussed in the previous section, and Taylor expanding the geodesics as given in Eq. (\ref{geodesic}), we can define the transformation between spacetime coordinates ($t,r,\theta,\phi$) and GN coordinate  ($t',\rho,\theta',\phi'$) upto $\mathcal{O}(\rho^2)$ as,
 \begin{eqnarray}
 && t'=t~;
 \nonumber
 \\ 
&& r = r_c + \rho~s_1(\theta) - \rho^2 s_2(\theta)~;
\nonumber
\\
&& \theta'= \theta.\ \ \phi' =\phi~. 
 \label{tranf1}
 \end{eqnarray}
 %Where the only non-zero component of the tangent vector to the geodesics is give by,
 where we define $s_1(\theta)=N^r \vline_{r=r_c} $, and 
 \begin{eqnarray}
  && s_2(\theta)=\frac{d^2 X^a}{d\rho^2} \vline_{\rho= 0}
  \nonumber
  \\
  &&=\Big(\frac{a^2 r_c-M r_c^2 +(M-r_c)a^2 \cos^2 \theta}{2(r_c^2+a^2 \cos^2 \theta)^2} \Big)~.
 \end{eqnarray}
 The above transformations upto $\mathcal{O}(\rho^2)$ will be sufficient for our subsequent discussion.
 Using the above transformation of coordinates, the induced metric near the timelike hypersurface under consideration will assume the GN form as Eq. (\ref{GN})
%\begin{eqnarray}
%&&ds^2= g_{tt}(\rho,\theta) dt^2 + g_{\rho \rho}(\rho,\theta) d\rho^2 + g_{t\phi}(\rho,\theta) dt d\phi \nonumber\\ &+& g_{\theta\theta}(\rho,\theta) d\theta^2  + g_{\phi \phi}(\rho,\theta) d\phi^2,\label{kerrmetric}
%\end{eqnarray}
where metric components upto $\mathcal{O}(\rho)$ are given by,
\begin{eqnarray}
g_{tt} & =& - \frac{\Delta_{r_c} -a^2 \sin^2\theta}{\Sigma_{r_c}}\nonumber\\ &-& 2 \rho~s_1(\theta)~\Big(\frac{M r^2_c -a^2 M\cos^2\theta}{\Sigma^2_{r_c}} \Big)~;
\nonumber
\\
g_{\rho \rho}&=& 1~;  
\nonumber
\\
g_{t\phi}&=& \frac{-4 a M r_c\sin^2\theta}{\Sigma_{r_c}}\nonumber\\ &+& 4 a M s_1(\theta) \rho ~\sin^2\theta \Big(\frac{r^2_c -a^2\cos^2\theta}{\Sigma^2{_{r_c}}}  \Big)~; 
\nonumber
\\ 
g_{\theta\theta} &=& \Sigma_{ r_c} +2 \rho~ r_c~ s_1(\theta)~; 
\nonumber
\\ 
g_{\phi\phi}&=& \sin^2\theta \Big(\frac{ (r_c^2 +a^2)^2 -a^2 \sin^2\theta \Delta_{r_c} }{\Sigma_{r_c}} \Big)
\nonumber
\\ 
&+& 2 \rho s_1(\theta) \sin^2\theta \Big[\Big(\frac{2a^2r_c+2 r^3_c + (a^2 M - a^2r_c) \sin^2\theta}{\Sigma_{r_c}} \Big) \nonumber\\  && 
-\Big(\frac{r_c(a^2+r_c^2)^2-a^2 r_c(a^2-2Mr_c+r^2_c)\sin^2\theta}{\Sigma^2{_{r_c}}} \Big) \Big]~.
\nonumber
\\
\label{Kcomp}
\end{eqnarray}
Here $\Sigma_{r_c}$ and $\Delta_{r_c}$ are the corresponding quantities defined at $r=r_c$.
These are given by,
\begin{eqnarray}
\Sigma_{r_c} &=& r_c^2 +a^2 \cos^2\theta~, 
\nonumber
\\
\Delta_{r_c} &=& r_c^2+a^2- 2M r_c~. 
\end{eqnarray}
Therefore, the above form (\ref{Kcomp}) in GN coordinate properly describes the spacetime geometry very near to the timelike surface situated at $\rho=0$ (where $r-r_c \approx \rho)$. Hence the components of the metric take the asymptotic form as given in (\ref{Kcomp}) near the aforementioned timelike boundary. Our subsequent analysis will be same as that of the Schwarzschild black hole background discussed above.

 \subsection{Symmetries near the surface}
 Unlike Schwarzschild black holes,  Kerr spacetime is endowed with  angular momentum which play characteristically different role in defining the properties of the associated symmetry algebra. 
 %So the appropriate boundary conditions has to be imposed on the variation of the metric components in order to determine the generators. 
 Main objective of the whole symmetry analysis program is to keep the asymptotic structure of the metric under investigation invariant under diffeomorphism transformation. Here also the following constraints on the behaviour of some of the metric components which are associated with the gauge condition, need to be maintained as before, and those are, 
\begin{equation}
\pounds_\zeta  g_{\rho \rho}= 0;  \  \pounds_\zeta  g_{\rho t}= 0; \  \pounds_\zeta  g_{\rho A}= 0~,
\label{boundary1}
\end{equation}
where $A$ stands for the transverse coordinates $\theta$ and $\phi$. Because of the inherent rotational angular momentum of the spacetime, the weak fall off conditions  of the remaining metric component in terms of $\rho$ coordinate are assumed to follow,
\begin{equation}
\pounds_\zeta  g_{tt} \approx \mathcal{O}(1); \ \pounds_\zeta  g_{tA} \approx \mathcal{O}(1); \  \pounds_\zeta  g_{AB} \approx \mathcal{O}(1)~. 
\label{boundary2}
\end{equation}
The fall-off conditions directly follow from the leading order behaviour of the metric coefficients $g_{tt}$, $g_{tA}$ and $g_{AB}$. The two categories of boundary conditions keep the asymptotic form of the metric intact near the timelike surface at $r=r_c$. So these are the asymptotic conditions imposed on the fluctuation of the metric coefficients (\ref{Kcomp}). Solving the three strong form of gauge conditions expressed in Eq. (\ref{boundary1}) we determine the general solutions of the symmetry vectors as,
\begin{eqnarray}
\zeta^{\rho} &=& T(t,x)~;
\nonumber
\\
\zeta^t &=& -\int \frac{\partial_t T}{g_{tt}} d\rho+ \int \frac{g^{tA}}{g_{tt}}[g_{tt} \partial_{A} T + g_{tA} \partial_t T)] d\rho\nonumber\\ & +& F(t,x)~;
\nonumber
\\
\zeta^{A} &=&  [-g^{tA}\int \partial_t T d\rho -g^{AB} \int \partial_{B} T d\rho]\nonumber\\ & +& R^{A}(t,x)~,
\label{Kerrsol}
\end{eqnarray}
where $x$ corresponds to the angular coordinates $\theta, \phi$. Here $T$ $F$, and $R^{A}$ are the unknown integration constants. These are the diffeomorphism vectors under which the form of the metric near the time like surface under consideration will remain invariant. 
Now further restrict on the above vectors under which the position of the surface does not change will lead to the following additional constraint $T(t,x)=0$. The fall-off conditions expressed in Eq. (\ref{boundary2}) have been observed to be  automatically satisfied by the solution space Eq. (\ref{Kerrsol}). Finally the non-vanishing components of the symmetry vectors $\zeta = \zeta^a \partial_a $ are given by,
\begin{eqnarray}
\zeta = {\cal F}(t,x) + {\cal R}(t,x) = F(t,x)\partial_t + R^{A}(t,x) \partial_A. \
\end{eqnarray}
In the above expression $R^{A}$ denotes the rotation parameter. For generality we assume all the differmorphism parameters to be function of all the coordinates namely $\zeta^\theta = R^\theta (t,\theta,\phi)$ and $\zeta^\phi= R^\phi (t,\theta,\phi)$. 
%The notation used in the subsequent discussion are as follows, $x^{A_1} = \theta,x^{A_2} = \phi$. $x^A$  is reserved for both the angular coordinate collectively. 
Following the conventional definition we describe the parameters $F$ as the generator of the `supertranslation' and $R^{A}$ as the `superrotation'. Important to note that for Kerr black hole background the supertranslation generator near the time-like surface turned out to be arbitrary function of both time and angular coordinates as opposed to that of the Schwarzschild black hole background discussed before.

Similar to Schwarzschild case the form of the vector field $\zeta$ is such that near surface $S$ the metric coefficients in kerr background keep their asymptotic form intact, while from the Lie variation of $g_{tt}$ one can show that the black hole mass $M$ has been changed to,
\begin{widetext}
\begin{eqnarray}
M \rightarrow M-\frac{(\Sigma_{r_c} - 2 r_c M)\Sigma_{r_c}\partial_t F -2 M a^2 r_c R^{\theta} \sin 2\theta-4 a M r_c \Sigma_{r_c}  \sin^2\theta (\partial_t R^{\phi} )}{2 r_c \Sigma_{r_c}};
\end{eqnarray}
and from the variation of the $g_{t\phi}$ component, $ a/\Sigma_{r_c}$ transform to,
\begin{eqnarray}
&&\frac{a}{\Sigma_{r_c} }
\rightarrow \frac{a}{\Sigma_{r_c} } (1+ \partial_{\phi} R^{\phi} -\partial_t F)\nonumber\\
&& +\frac{4 a M r_c(\Sigma_{r_c}+a^2  \cos\theta) R^{\theta} \sin2 \theta -(\Sigma_{r_c} -2 M r_c) \Sigma^2_{r_c} \partial_{\phi} F}{4 M r_c \Sigma^2_{r_c} \sin^2\theta} -\frac{\Big((r_c^2 +a^2)^2 -a^2 \sin^2 \theta \Delta_{r_c}\Big) \partial_t R^{\phi}}{4 M r_c \Sigma^2_{r_c}}~.
\end{eqnarray}
\end{widetext}
Following the same procedure discussed for the Schwarzschild black hole, we now explore the bracket algebra among the three symmetry generators  $\chi^a=(F,0,0)$ and $\xi^a=(0,0,R^{A})$. 
%The vectors are chosen in such a way that $\chi^a$ has only time component, $\chi^t= F(t,x^A)$ which is non-zero and the other vector $ \xi^a$ has only non-vanishing angular component, $\xi^{A_q}=R^{A_q}(t,x^A)$. 
The generators are expressed in terms of following Fourier expansion,
\begin{eqnarray}
{\cal F}(t,x) &=& \sum_{k_A} \int_{-\infty}^{\infty} d\omega~ \alpha(\omega, k_A) e^{- i \omega t + i  k_A x^{A}}\partial_t\nonumber\\
%\int_{-\infty}^{\infty} d\omega~ \{\alpha(\omega) ~e^{-i \omega t}\}~ \{\Sigma_{k} \alpha_k ~  e^{i\Sigma_{A_q}  ~k ~x^{A_{q}}}\}\nonumber\\
&=& \sum_{k_A}\int_{-\infty}^{\infty}  d \omega ~\alpha(\omega, k_A) ~\bar{F}_{k}(\omega,t,x); \label{mode}\\
{\cal R}(t,x) &=& \sum_{l_A} \int_{-\infty}^{\infty} d\omega~ \bar{\alpha}^{A}(\omega, l_A)~   e^{-i \omega t +  i   l_A  x^{A}} \partial_A\nonumber\\
&=& \sum_{l_A} \int_{-\infty}^{\infty} d \omega~ \bar{\alpha}^{A}(\omega, l_A)  \bar{R}^{l}_{A}(\omega,t,x). \label{mode1}
\end{eqnarray}
In the above expression $ \alpha(\omega)$ and $\bar{\alpha}^{A}(\omega)$ represent the Fourier modes in frequency domain of the functions $F(t,x)$ and $R^{A}(t,x)$ respectively. The argument $x$ corresponds to angular coordinates $(\theta,\phi)$. $\bar{F}_k(\omega,t,x)$ and $\bar{R}^{l}_A(\omega,t,x)$ are identified as basis generators among which the commutator algebra will be as follows,
\begin{widetext}
By using Eq. (\ref{Lie}), the required bracket for the `supertranslation' vector with itself 
%\begin{eqnarray}
%[\chi_1,\chi_2]^t = [F_1,F_2]= F_1 \partial_t F_2- (1\leftrightarrow 2).
%\end{eqnarray}
can be calculated as,
\begin{eqnarray}
&&[F_1,F_2]= \sum_{k_A,l_A}\int_{-\infty}^{\infty} d\omega_1 \ d\omega_2~  \alpha(\omega_1,k_A)\alpha(\omega_2,l_A)  [\bar{F}_{k}(\omega_1,t,x),\bar{F}_{l}(\omega_2,t,x)].\label{f1}
\end{eqnarray}
After expanding the bracket we obtain,
\begin{eqnarray}
&&F_1 \partial_t F_2- (1\leftrightarrow 2)= \sum_{k_A,l_A}\int_{-\infty}^{\infty} d\omega_1~ d\omega_2 ~ [-i(\omega_2 -\omega_1)]~ \alpha(\omega_1,k_A) \alpha(\omega_2,l_A)  \bar{F}_{k+l}(\omega_1+\omega_2,t,x),\label{f2}
\end{eqnarray}
Comparing expression (\ref{f1}) with (\ref{f2}) we get,
\begin{eqnarray}
i[\bar{F}_{k}(\omega_1,t,x),{\bar F}_{l}(\omega_2,t,x)]= (\omega_2 -\omega_1)~ \bar{F}_{k+l}(\omega_1 +\omega_2,t,x). \label{Fk}
\end{eqnarray}
By using  the similar manner we can calculate the commutator algebra between `supertranslation' and `superrotation' parameters as 
\begin{eqnarray}
&& i[\bar{F}_{k}(\omega_1,t,x),\bar{R}_{A}^{l}(\omega_2,t,x)] 
=\omega_2 \bar{R}_{A}^{k+l}(\omega_1 +\omega_2,t,x)~-~ k_A  \bar{F}_{k+l}(\omega_1+\omega_2,t,x)~;
\end{eqnarray}
%\begin{eqnarray}
%&& i[\bar{F}_{k_A}(\omega_1,t,x^A),\bar{R}^{A}_{l_A}(\omega_2,t,x^A)]^t 
%=- k ~ \bar{F}_{k+l}(\omega_1+\omega_2,t,x^A)~;
%\nonumber\\ 
%\label{FRa}
%&& i[\bar{F}_k(\omega_1,t,x^A),\bar{R}^{A_{q}}_l(\omega_2,t,x^A)]^{A'_q} = \omega_2~ \delta_{qq'}~\bar{R}^{A'_{q}}_{k+l}(\omega_1 +\omega_2,t,x^A)~.
%\label{FrA}
%\end{eqnarray}
%\end{eqnarray}
The Lie bracket for the `superrotation' diffeomorphism vector with itself will take the following form,
\begin{eqnarray}
 i[\bar{R}_{A}^{k}(\omega_1,t,x)],\bar{R}_{B}^{l}(\omega_2,t,x)] 
 = l_A~  \bar{R}_{B}^{k+l}(\omega_1 +\omega_2,t,x)
 - k_B~  \bar{R}_{A}^{k+l}(\omega_1 +\omega_2,t,x)~.
 \label{FrR}
\end{eqnarray}
\end{widetext}
Let us point out important difference between the results obtained for the time like surface embedded in Kerr spacetime from that of the null surface \cite{Maitra:2018saa}.
For the null surface case, the notable difference is that the general `super-rotation' generators is time independent as opposed to the present case. Therefore, this fact leads to  difference result for the $[\mathcal{R}(t,x), \mathcal{F}(t,x)]$ commutator. On the null surface the aforementioned commutator  generates super-translation only where as on the time like surface it generates both `super-rotation' and `super-translation' depending upon the parameter values of the transformation. 
%%%%%%%%%%%%%%%%%%%%%%%%%%%%%%%%%%%%%%%%%%%%%%%%%%%%%%%%%%%%%%%

\subsection{ Charges $Q$ on the surface}
%After having algebra among symmetry parameters our next move will be to get the expression of conserved charges conjugate to the aforesaid symmetry generators. The reason behind formulating conserved quantities have been discussed already for the Schwarzschild case in the last section. 
%It has been shown \cite{} that the thermodynamical properties of the rotating black hole especially horizon entropy can be determined from the Noether charges associated to symmetries. 
%Hence we want to explore the conserved charges associated with the underlined symmetries near the given hypersurface in Kerr spacetime background. Later these charges may be helpful to determine the thermodynamics properties of the timelike hypersurface. We will derive charges for Einstein gravity which will be off-shell.

Associated with the symmetries discussed in the previous section, we now explicitly compute the charges $Q$ defined on the two dimensional $r=$ constant, $t =$ constant surface which is essentially a two sphere. The sphere at a particular instant of time is characterized by a spacelike and a timelike normal vectors defined as $M^a$ and $n^a$ respectively. 
For the metric Eq. (\ref{Kcomp}), the unit spacelike normal vector $M^a$ assumes $M^a= (0,1,0,0)$
%\begin{eqnarray}
%M^{\rho} =\frac{1}{s(\theta)}\sqrt{\Delta_{\rho}/ \Sigma_{\rho}},
%\end{eqnarray}
and the non-zero components of the unit timelike normal vector $n^a$ are:
\begin{eqnarray}
n^t&=& -\sqrt{\frac{\alpha(\rho,\theta)}{\Sigma(\rho,\theta) \Delta(\rho,\theta)}} \\
~~~~n^{\phi}&= &
-\frac{2 a M (r_c + \rho~s_1-\rho^2 s_2)}{\sqrt{\alpha(\rho,\theta)}}~,
\end{eqnarray}
\begin{widetext}
where we have defined the following quantities in GN coordinate,
\begin{eqnarray}
\Sigma(\rho,\theta) &=& (r_c + \rho~s_1-\rho^2 s_2)^2 +a^2 \cos^2\theta, \nonumber
\\
\Delta(\rho,\theta) &=& (r_c + \rho~s_1-\rho^2 s_2)^2 + a^2- 2M (r_c + \rho~s_1-\rho^2 s_2) , \nonumber
\\
\alpha(\rho,\theta) &=& ((r_c + \rho~s_1-\rho^2 s_2)^2 +a^2)^2 -a^2 \sin^2\theta \Delta(\rho,\theta), \nonumber
%\beta(\rho,\theta)&=& 16 a^2 M^2(r_c + \rho~s_1-\rho^2 s_2)^2 \sin^2\theta
%- \Big( 2 M (r_c - \rho~s_1-\rho^2 s_2)  -(r_c + \rho~s_1-\rho^2 s_2)^2 - a^2 \cos^2\theta \Big) %\alpha(\rho,\theta).\nonumber\\ \label{beta}
\end{eqnarray}
Following the same procedure as discussed for the Schwarzschild case, the area element $d\Sigma^{ab}$ survives with two non vanishing components: $d\Sigma^{t\rho}$ and $d\Sigma^{\rho \phi}$.
With all these ingredients the diffeomorphism charges can be expressed near the surface $\rho=0$, as follows,
\begin{eqnarray}
Q[\zeta]= Q[\mathcal{F}]+ Q[\mathcal{R}] ~,
\end{eqnarray}
%So the conserved charges originating due to supertranslation and superrotation diffeomorphism symmetries can be expressed separately as follows,
%Expressions for both the parts come out to be
The first term is `super-translation'  charge with the following explicit form, 
\begin{eqnarray}
Q[\mathcal{F}]=  -\frac{1}{16 \pi G}  \int d^2 x  \Big[ 2 M (\sin\theta) \mathcal{F}(t,x)\frac{ (a^2 +r_c^2) (r_c^4-a^4 \cos^4\theta)}{(r_c^2+a^2 \cos^2\theta)^3} \Big]~,\label{QFF}
\end{eqnarray}
%\begin{eqnarray}
%&& Q[F]=  \frac{1}{4 \pi G}  \int d^2 x \Big[F(t,x^A) 
%   {\frac{ M (\sin\theta) \sqrt{\Delta_{r_c}}  (r^2_c -a^2 \cos^2\theta)}{\Sigma^2_{r_c}}}   %\frac{(a^2+r_c^2)^2-(a^2+6 M r_c +r_c^2)a^2 \sin^2\theta}{\sqrt{\beta_{r_c}}} \Big]~,\nonumber\\
%  \label{Qc}
%\end{eqnarray}
%and
the second term is attributed to the charge associated with the `superrotation',
\begin{eqnarray}
 Q[\mathcal{R}]= \frac{1}{16 \pi G}\int d^2 x \Big[ 2 a M (\sin\theta) \mathcal{R} (t,x) \frac{(3 r_c^6 +a^2 r_c^4 + 4 a^2 r_c^4 \cos^2\theta- a^4 (a^2-r_c^2) \cos^4\theta ) }{(r_c^2+a^2 \cos^2\theta)^3} \Big]~.
 \label{QR}
\end{eqnarray}
\end{widetext}
Now we are in a position to inspect the physical interpretation of the `supertranslation' and `superrotation' charges. For simplest choice,  $F(t,x)=1$ and $R^A(t,x)= 1$ the results (\ref{QFF}) and (\ref{QR}) yield the followings form of the charges,
\begin{eqnarray}
Q[F(t,x)=1]= \frac{M}{2 G},~~~~Q[R^A(t,x)=1]= -\frac{M a}{G};
\end{eqnarray}
which can be identified as the Komar conserved quantities \cite{Komar:1958wp,Banerjee:2010yd,Banerjee:2010ye} defined on the timelike surface at a particular instant of time in the Kerr black hole spacetime.  However, with the given diffeomorphism vectors, we have not found any obvious thermodynamic interpretation as has been found for the Schwarzchild case. However, if we consider
a slightly more general diffeomorphism vectors $\zeta = \partial_t + \Omega_c\partial_{\phi}$, where   
\begin{eqnarray}
\Omega_c &=& -\frac{g_{t\phi}}{g_{\phi\phi}}\vline_{r=r_c} \\\nonumber
&=&\frac{2 M a r_c}{(r_c^2+a^2) (r_c^2+a^2 \cos^2\theta)+2 a M r_c \sin^2\theta}~,\label{omegaC}
\end{eqnarray} 
is the the angular velocity on our timelike surface.
In that case we have 

%Moreover for the choice $F(t,x)=1$ and $R^A(t,x)= \Omega_H$, where the angular velocity at the horizon is given by
%\begin{eqnarray}
%\Omega_H= \frac{a}{2 M (M+ \sqrt{M^2-a^2})}~,
%\label{omegaH}
%\end{eqnarray}
%which corresponds to the timelike Killing vector $\zeta^a= (1,0,0,\Omega_H)$ one obtains,
%\begin{eqnarray}
% && Q_{\zeta}= Q[\bar{F}=1] + Q[\bar{R}^A=\Omega_H]\nonumber
% \\
%&&= \frac{M^2 - a^2 + M \sqrt{M^2-a^2}}{2G ( M + \sqrt{M^2-a^2})}~.
%\label{QH}
%\end{eqnarray}

\begin{widetext}
\begin{eqnarray}
&& Q_{\zeta} = Q[F(t,x)=1] + Q[R^A(t,x)=\Omega_c]\nonumber
\\
&& =\frac{a^2-2 M r_c+r_c^2}{2 a G} \tan^{-1}{(\frac{a}{r_c})} -\frac{(a^2+ r_c^2)(a^2 r_c + a^2 M + r_c^3-3 M r_c^2)}{2 a G  \sqrt{r_c} \sqrt{(a^2 + r_c^2-2 M r_c)(r_c^3+2 M a^2+a^2 r_c)}} \tan^{-1}{[\frac{a \sqrt{a^2-2 M r_c +r_c^2}}{\sqrt{r_c(r_c^3+2 M a^2 +a^2 r_c )}}]}. \nonumber
\\ \label{Q'}
\end{eqnarray}
%One can easily check that in the limit $ r_c \rightarrow r_H $ the angular velocity (\ref{omegaC}) at the distance $r_c$  reduces to the horizon angular velocity (\ref{omegaH}) and the expression of the charge (\ref{Q'}) becomes (\ref{QH}). But in this case, using the available expression for Tolman temperature (see \cite{Santiago:2018lcy} for instance) in the case of Kerr spacetime, we could not able to express none of (\ref{QH}) and (\ref{Q'}) as the form (\ref{B5}). It is mentioned in \cite{Santiago:2018lcy} that the proper way of defining the Tolman temperature for stationary case is not known, except some proposals. It may happen that due to non-availability of this suitable expression we failed to write the above in the form (\ref{B5}) and hence we leave the interpretation of the value of our charge for future. 
To find similar relation obtained in Eq. (\ref{B55}), following the reference \cite{Santiago:2018lcy}, we generalize the Tolman relation for stationary spacetime as $T(r) \sqrt{-g_{ab}\zeta^a \zeta^b} =T_0$ where $\zeta^a$ is the timelike vector, constructed out from the linear combination of two existing Killing vectors $(\partial_t)^a$ and $(\partial_\phi)^a$ of the spacetime.  Now, in this case the form of $\zeta^a$ is taken to be $\zeta^a=(1,0,0,\Omega_c)$ and we define
\begin{eqnarray}
|K|= \sqrt{-(g_{tt} +2 \Omega_c g_{t\phi} +\Omega^2_c g_{\phi\phi})}~.
\end{eqnarray}
Then the Tolman temperature gradient leads to the gravitational acceleration along the normal to the surface as follows,
 \begin{eqnarray}
 g=- \hat{n}^a \frac{\nabla_a {T}}{T}= - \sqrt{g^{rr}} \frac{\partial_r |K|^2 }{|K|^2},
 \end{eqnarray}
 where  $\hat{n}^a$ is the normal to the $r=$constant surface given by, $(0, 1/\sqrt{g_{rr}},0,0)$.
 Associated expression for the charge (\ref{Q'}), in differential form, will then take the identical form as,
 \begin{eqnarray}
 \delta Q_{\zeta}=- \frac{1}{2 \pi}  \left(\frac{\delta{A}_c}{4 G} \right) \left(\frac{T_{0}}{T} \right) \left(\frac{\hat{n}^a\nabla_a T}{T} \right)\vline_{r=r_c}~.\label{Q'Tolman}
 \end{eqnarray}
 Area element $\delta{A}_c$ is given by  $\sqrt{\alpha_c} \sin\theta~d\theta d\phi$. It is now easy to check that after integration over transverse coordinates the expression (\ref{Q'Tolman}) reduces to (\ref{Q'}).

%\begin{eqnarray}
%&& Q[R^{A}]= \frac{1}{2\pi G}\int d^2 x \Big[R^{A}(t,x^A)   \frac{a M (\sin^3 \theta) %\sqrt{\Delta_{r_c}}\{ -\alpha_{r_c} (r^2_c-a^2 \cos^2\theta)- r_c 
%(\Sigma_{r_c} \gamma_{r_c}-2r_c \alpha_{r_c})\}}{\Sigma^2_{r_c} \sqrt{\beta_{r_c}}}\Big],
%\label{QRc}
%\end{eqnarray}
%where we have defined,
%\begin{eqnarray}
%&&\gamma_{r_c} = 4 a^2 r_c+4 r^3_c +2 a^2(M-r_c) \sin^2\theta~;
%\nonumber\\
%\Sigma_{r_c}= \Sigma_{\rho}\vline_{\rho=0};  \ \ \Delta_{r_c}= \Delta_{\rho}\vline_{\rho=0}\nonumber\\
%&&\alpha_{r_c}=\alpha \vline_{\rho=0}, \ \  \beta_{r_c}= \beta\vline_{\rho=0}~.
%\end{eqnarray}

The charges can now be expressed in terms of the mode function of the symmetry generators as,
\begin{eqnarray}
 Q[F(t,x)] &=& \sum_k \int_{-\infty}^{\infty}  d \omega~ \alpha_k(\omega)~ Q[\bar{F}_k(\omega,t,x)]~;
 \nonumber \\
Q[{\cal R}(t,x)]&=& \sum_l \int_{-\infty}^{\infty} d \omega ~\bar{\alpha}_l^{A}(\omega)~ Q_A[\bar{R}^l(\omega,t,x)]~,\nonumber\\ \label{KerrQ}
\end{eqnarray}
where each mode can be written as follows,
%$Q[\bar{F}_k(\omega,t,x^A)]$ 
\begin{eqnarray}
Q[\bar{F}_k(\omega,t,x)]=  -\frac{1}{16 \pi G}  \int d^2 x  \Big[2 M (\sin\theta) \bar{F}_k(\omega,t,x)\frac{ (a^2 +r_c^2) (r_c^4-a^4 \cos^4\theta)}{(r_c^2+a^2 \cos^2\theta)^3} \Big]~,
\end{eqnarray}
%\begin{eqnarray}
and
\begin{eqnarray}
 Q_A[\bar{R}^l(\omega,t,x)]= \frac{1}{16 \pi G}\int d^2 x \Big[ 2 a M (\sin\theta) \bar{R}_{A}^l(\omega,t,x)\frac{(3 r_c^6 +a^2 r_c^4 + 4 a^2 r_c^4 \cos^2\theta- a^4 (a^2-r_c^2) \cos^4\theta ) }{(r_c^2+a^2 \cos^2\theta)^3} \Big] ~.
 \end{eqnarray}
%\begin{widetext}
Then  using Eq. (\ref{ch}) the algebra between modes of charges are computed as follows:
 \begin{eqnarray}
 %\begin{center}
 &&i[Q[\bar{F}_k(\omega_1,t,x)],Q[\bar{F}_l(\omega_2,t,x)]] =  (\omega_2 -\omega_1) 
 Q[\bar{F}_{k+l}(\omega_1 +\omega_2,t,x)]~;
 \nonumber
 \\
  && i[Q[\bar{F}_k(\omega_1,t,x)],Q_A[\bar{R}^l(\omega_2,t,x)]] = - k_A~ Q[ \bar{F}_{k+l}(\omega_1 +\omega_2,t,x)] ~+~ \omega_2 \ Q_A[\bar{R}^{k+l}(\omega_1 + \omega_2,t,x)];
  \nonumber
  \\
 && i [Q_A[\bar{R}^k(\omega_1,t,x)], Q_B[\bar{R}^l(\omega_2,t,x)]]= l_A~ Q_B[ \bar{R}^{k+l}(\omega_1 +\omega_2,t,x)]- k_B~Q_A[ \bar{R}^{k+l}(\omega_1 +\omega_2,t,x)]~.
 \label{Qbracket}
  \end{eqnarray}
  \end{widetext}
   It is clear from the above Eq. (\ref{Qbracket}) that the symmetry bracket among the charges are isomorphic to that among diffeomorphism vectors. Hence we can conclude that the components of the diffeomorphism symmetry generators together form a closed algebra which is slightly different from the standard near horizon BMS algebra.
  
%%%%%%%%%%%%%%%%%%%%%%%%%%%%%%%%%%%%%%%%%%%%%%%%%%%%%%%%%%%%%%%%%%55
  \section{conclusion}
 In this paper we have studied in detail the symmetries of a timelike hypersurface positioned at any arbitrary radial coordinate embedded in black hole spacetime. We have considered two different black hole spacetime, where symmetries have been identified by considering the class of diffeomorphisms which keeps the form of the metric near the time like surface invariant. In the present analysis the obtained bracket algebra of the asymptotic symmetry group near the timelike surface has similarities with those of the black hole horizon and null infinity in Schwarzschild background but for kerr spacetime there is significant difference between the algebra near the timelke surface and those near actual physical boundaries of spacetime. %The symmetries near the $r$-constant hyper-surface have been observed to follow quite different algebra as that of the asymptotic infinity or near black hole horizon. 
 %The symmetry algebra turned out to be little bit different compared to that of the surfaces located at the horizon and at the asymptotic infinity of the black hole. 
 This is because in Schwarzschld spacetime  the symmetry generator defined on the timelike surface turns out to be dependent only on time, but the bracket algebra of the `supertranslation' generator is similar to  that obtained for a generic null surface . Whereas for Kerr case `superrotation` parameter $\mathcal{R}(t,x)$ is time dependent, therefore notable differences has generated for the $[\mathcal{R}(t,x),\mathcal{F}(t,x)]$ commutator compared to null boundaries. The associated  charges have also been computed and shown to follow the same algebra as differomorphism transformation vectors. We have also established an interesting connection between the charges with the local heat content on the surface under the study.  
 
 The timelike surface divides the space  into two complementary regions which are  causally disconnected at any instant of time. Therefore, the symmetries near the surface could play interesting role in understanding the entanglement phenomena in quantum theory. Usually all the symmetry analysis have been studied so far in the asymptotic infinity or near the horizon keeping in mind the causal physical phenomena such as scatting of massive or massless particles. Role of symmetries in acausal phenomena such as entanglement has not been discussed in the literature. Our present analysis of symmetries near timelike surfaces could be useful to understand this acausal phenomena which will be taken up in future publication.     
 %%%%%%%%%%%%%%%%%%%%%%%%%%%%%%%%%%%%%%%%%%%%%%%
 \vskip 3mm
 \noindent
 {\bf Acknowledgement:}
 {\scshape{This work is dedicated to those who are helping us to fight against COVID-19 across the globe.}}
 
 %%%%%%%%%%%%%%%%%%%%%%%%%%%%%%%%%%%
 
 \appendix
  \section{Local acceleration of the observer at $r=r_c$}\label{App1}
  The effective metric, from (\ref{B2}), for an observer moving along $\rho$ in the vicinity of $r=r_c$ surface is given by 
  \begin{eqnarray}
  ds^2 &=& -[f_1(r_c) + f_1'(r_c)\sqrt{1/f_2(r_c)}~\rho]~dt^2 + d\rho^2 .
  \label{app1} 
  \end{eqnarray}
  The above one is in Rindler form and more conveniently can be expressed by the transformation $f_1(r_c) + f_1'(r_c)\sqrt{1/f_2(r_c)}~\rho=x$. Then (\ref{app1}) reduces to,
  \begin{eqnarray}
ds^2= -x dt^2 + \frac{f_2 (r_c)}{f'^2_1(r_c)} dx^2~.
\label{app2} 
  \end{eqnarray}
The metric (\ref{app2}) can be expressed in Minkowski form  $ds^2= -dT^2 + dX^2$, by having following transformation of coordinates,
  \begin{eqnarray}
  X= \frac{\sqrt{x}}{a} \cosh (a t)~,~~~ T= \frac{\sqrt{x}}{a} \sinh (a t)~,
  \label{AppB1}
\end{eqnarray} 
where 
\begin{equation}
a(x)=\frac{f'_1(r_c)}{2 \sqrt{f_2(r_c) x}}~.
\label{AppB2}
\end{equation} 
The coordinates $(t,x)$, adapted to the uniformly accelerated motion, is known as Rindler coordinates. Below we will show that $a$ is identified to be the local acceleration of an observer in Rindler frame.

  To show this let us first calculate the proper acceleration this observer. For that we consider any $x=$ constant trajectory and then for this value of $x$ the coordinate time $t$ is identified as the proper time $\tau$.  Then the magnitude of the proper acceleration, defined as $a_{prop}=\sqrt{a^i_{prop}a_{i_{prop}}}$ with $a^i_{prop} = dX^a/d\tau$, from (\ref{AppB1}) is obtained as, 
  \begin{eqnarray}
  a_{prop} = a(x) \sqrt{x} \vline_{x=const}~. %= \frac{f'_1(r_c)}{2 \sqrt{f_2 (r_c)}}\vline_{x=const}.
  \end{eqnarray}
Therefore the local acceleration is given by $ a(x) =a_{prop} / \sqrt{x}$. At $r=r_c$, from (\ref{AppB2}), this is given by
\begin{eqnarray}
  g(r_c) = a(x=x_c) = \frac{a_{prop}}{\sqrt{x_c}}= \frac{f'_1(r_c)}{2 \sqrt{f_2 (r_c) f_1(r_c)}}. 
\end{eqnarray}    
  
 % Hence for the motion along the path having $x=1$ , observer will feel acceleration $a= a_{prop}$. But for any other $x=$constant path,  acceleration will be $ a(x) =a_{prop} / \sqrt{x}=\frac{f'_1(r_c)}{2 \sqrt{f_2(r_c) x}}$.
  
 %  So in GNC local acceleration at $r=r_c$ is given by,

%%%%%%%%%%%%%%%%%%%%%%%%%%%%%%%%%%%%%%%%

% \end{eqnarray}

\begin{thebibliography}{99}

\bibitem{Iyer:1994ys} 
  V.~Iyer and R.~M.~Wald,
  Some properties of Noether charge and a proposal for dynamical black hole entropy,
  Phys.\ Rev.\ D {\bf 50}, 846 (1994).
  %doi:10.1103/PhysRevD.50.846
  [gr-qc/9403028].
  %%CITATION = doi:10.1103/PhysRevD.50.846;%%
  
  %\cite{Wald:1993nt}
 \bibitem{Wald:1993nt}
 R.~M.~Wald,
 ``Black hole entropy is the Noether charge,''
 Phys. Rev. D \textbf{48}, no.8, 3427-3431 (1993)
 %doi:10.1103/PhysRevD.48.R3427
[arXiv:gr-qc/9307038 [gr-qc]].

 %\cite{Wald:1999w
  \bibitem{Wald:1999wa}
R.~M.~Wald and A.~Zoupas,
``A General definition of 'conserved quantities' in general relativity and other theories of gravity,''
Phys. Rev. D \textbf{61}, 084027 (2000)
doi:10.1103/PhysRevD.61.084027
%[arXiv:gr-qc/9911095 [gr-qc]].
%378 citations counted in INSPIRE as of 05 May 2021
  

%\cite{Bekenstein:1973ur}
\bibitem{Bekenstein:1973ur}
J.~D.~Bekenstein,
``Black holes and entropy,''
Phys. Rev. D \textbf{7}, 2333-2346 (1973)
%doi:10.1103/PhysRevD.7.2333
%4880 citations counted in INSPIRE as of 08 Oct 2020

%\cite{Bekenstein:2020fbz}
%\cite{Bekenstein:1974ax}
\bibitem{Bekenstein:1974ax}
J.~D.~Bekenstein,
``Generalized second law of thermodynamics in black hole physics,''
Phys. Rev. D \textbf{9}, 3292-3300 (1974)
%doi:10.1103/PhysRevD.9.3292
%1667 citations counted in INSPIRE as of 08 Oct 2020

%\cite{Bekenstein:2020eas}
%\cite{Bekenstein:1972tm}
\bibitem{Bekenstein:1972tm}
J.~D.~Bekenstein,
``Black holes and the second law,''
Lett. Nuovo Cim. \textbf{4}, 737-740 (1972)
%doi:10.1007/BF02757029
%1053 citations counted in INSPIRE as of 08 Oct 2020
  
  %\cite{Bardeen:1973gs}
\bibitem{Bardeen:1973gs}
J.~M.~Bardeen, B.~Carter and S.~W.~Hawking,
``The Four laws of black hole mechanics,''
Commun. Math. Phys. \textbf{31}, 161-170 (1973)
%doi:10.1007/BF01645742

    \bibitem{Hawking:1974sw} 
  S.~W.~Hawking,
  ``Particle Creation by Black Holes,''
  Commun.\ Math.\ Phys.\  {\bf 43}, 199 (1975)
  Erratum: [Commun.\ Math.\ Phys.\  {\bf 46}, 206 (1976)].
 %5 doi:10.1007/BF02345020
  %%CITATION = doi:10.1007/BF02345020;%%
  
    \bibitem{Hawking:1976de} 
  S.~W.~Hawking,
  ``Black Holes and Thermodynamics,''
  Phys.\ Rev.\ D {\bf 13}, 191 (1976).
  %6doi:10.1103/PhysRevD.13.191
  %%CITATION = doi:10.1103/PhysRevD.13.191;%%
  
    \bibitem{Gibbons:1977mu} 
  G.~W.~Gibbons and S.~W.~Hawking,
  ``Cosmological Event Horizons, Thermodynamics, and Particle Creation,''
  Phys.\ Rev.\ D {\bf 15}, 2738 (1977).
  %7doi:10.1103/PhysRevD.15.2738
  %%CITATION = doi:10.1103/PhysRevD.15.2738;%%
  
  
  
\bibitem{Hawking:1982dh} 
  S.~W.~Hawking and D.~N.~Page,
  ``Thermodynamics of Black Holes in anti-De Sitter Space,''
  Commun.\ Math.\ Phys.\  {\bf 87}, 577 (1983).
  %8doi:10.1007/BF01208266
  %%CITATION = doi:10.1007/BF01208266;%%

  
  \bibitem{Strominger:1997eq} 
  A.~Strominger,
  ``Black hole entropy from near horizon microstates'',
  J. High Energy Phys. 02 (1998) 009.
  %JHEP {\bf 9802}, 009 (1998)
  %doi:10.1088/1126-6708/1998/02/009
  [hep-th/9712251].
  %%CITATION = doi:10.1088/1126-6708/1998/02/009;%%  
  
\bibitem{Carlip:1999cy} 
  S.~Carlip,
  ``Entropy from conformal field theory at Killing horizons'',
  Classical  Quantum  Gravity   {\bf 16}, 3327 (1999)
  %doi:10.1088/0264-9381/16/10/322
  [gr-qc/9906126].
  %%CITATION = doi:10.1088/0264-9381/16/10/322;%%  

\bibitem{Carlip:1998wz}
  S.~Carlip,
  ``Black hole entropy from conformal field theory in any dimension'',
  Phys.\ Rev.\ Lett.\  {\bf 82}, 2828 (1999).
  %doi:10.1103/PhysRevLett.82.2828
  [hep-th/9812013].
  %%CITATION = doi:10.1103/PhysRevLett.82.2828;%% 
  
  \bibitem{Majhi:2011ws} 
  B.~R.~Majhi and T.~Padmanabhan,
  Noether Current, Horizon Virasoro Algebra and Entropy,
  Phys.\ Rev.\ D {\bf 85}, 084040 (2012).
  %doi:10.1103/PhysRevD.85.084040
  [arXiv:1111.1809 [gr-qc]].
  %%CITATION = doi:10.1103/PhysRevD.85.084040;%% 
  
\bibitem{Majhi:2012tf} 
  B.~R.~Majhi and T.~Padmanabhan,
  ``Noether current from the surface term of gravitational action, Virasoro algebra and horizon entropy'',
  Phys.\ Rev.\ D {\bf 86}, 101501 (2012).
  %doi:10.1103/PhysRevD.86.101501
  [arXiv:1204.1422 [gr-qc]].
  %%CITATION = doi:10.1103/PhysRevD.86.101501;%%
  
  \bibitem{Majhi:2012nq} 
  B.~R.~Majhi,
  ``Noether current of the surface term of Einstein-Hilbert action, Virasoro algebra and entropy'',
  Adv.\ High Energy Phys.\  {\bf 2013}, 386342 (2013).
  %doi:10.1155/2013/386342
  [arXiv:1210.6736 [gr-qc]].
  %%CITATION = doi:10.1155/2013/386342;%% 
  
\bibitem{Majhi:2013lba} 
  B.~R.~Majhi and S.~Chakraborty,
  ``Anomalous effective action, Noether current, Virasoro algebra and Horizon entropy'',
  Eur.\ Phys.\ J.\ C {\bf 74}, 2867 (2014).
  %doi:10.1140/epjc/s10052-014-2867-6
  [arXiv:1311.1324 [gr-qc]].
  %%CITATION = doi:10.1140/epjc/s10052-014-2867-6;%%  

\bibitem{Majhi:2014lka} 
  B.~R.~Majhi,
  ``Conformal Transformation, Near Horizon Symmetry, Virasoro Algebra and Entropy'',
  Phys.\ Rev.\ D {\bf 90}, 044020 (2014).
  %doi:10.1103/PhysRevD.90.044020
  [arXiv:1404.6930 [gr-qc]].
  %%CITATION = doi:10.1103/PhysRevD.90.044020;%%  
  
\bibitem{Majhi:2015tpa} 
  B.~R.~Majhi,
  ``Near horizon hidden symmetry and entropy of Sultana-Dyer black hole: A time dependent case'',
  Phys.\ Rev.\ D {\bf 92}, 064026 (2015).
  %doi:10.1103/PhysRevD.92.064026
  
 \bibitem{Majhi:2017fua}
B.~R.~Majhi,
``Noncommutativity in near horizon symmetries in gravity,''
Phys. Rev. D \textbf{95}, no.4, 044020 (2017)
%doi:10.1103/PhysRevD.95.044020
[arXiv:1701.07952 [gr-qc]].

\bibitem{Bhattacharya:2018epn}
K.~Bhattacharya and B.~R.~Majhi,
``Noncommutative Heisenberg algebra in the neighbourhood of a generic null surface,''
Nucl. Phys. B \textbf{934}, 557-577 (2018)
%doi:10.1016/j.nuclphysb.2018.07.025
[arXiv:1802.02862 [gr-qc]].
  
  \bibitem{Cardy:1986ie}
J.~L.~Cardy,
``Operator Content of Two-Dimensional Conformally Invariant Theories,''
Nucl.\ Phys.\ B \textbf{270}, 186-204 (1986)
doi:10.1016/0550-3213(86)90552-3

\bibitem{Brown:1986nw} 
  J.~D.~Brown and M.~Henneaux,
  ``Central Charges in the Canonical Realization of Asymptotic Symmetries: An Example from Three-Dimensional Gravity'',
  Commun.\ Math.\ Phys.\  {\bf 104}, 207 (1986).
  %doi:10.1007/BF01211590
  %%CITATION = doi:10.1007/BF01211590;%% 
  
  \bibitem{Parattu:2013gwa}
K.~Parattu, B.~R.~Majhi and T.~Padmanabhan,
``Structure of the gravitational action and its relation with horizon thermodynamics and emergent gravity paradigm,''
Phys. Rev. D \textbf{87}, no.12, 124011 (2013)
%doi:10.1103/PhysRevD.87.124011
[arXiv:1303.1535 [gr-qc]].
  
  \bibitem{Chakraborty:2015aja}
   S.~Chakraborty, K.~Parattu and T.~Padmanabhan,
  ``Gravitational field equations near an arbitrary null surface expressed as a thermodynamic identity'',
 J. High Energy Physics. {\bf 10}, (2015) 097;
  [arXiv:1505.05297 [gr-qc]];
      
\bibitem{Dey:2020tkj}
S.~Dey and B.~R.~Majhi,
``A covariant approach to thermodynamic structure of a generic null surface: old wine in a new bottle,''
[arXiv:2009.08221 [gr-qc]].      
 
 \bibitem{Weinberg:1965nx} 
  S.~Weinberg,
  ``Infrared photons and gravitons,''
  Phys.\ Rev.\  {\bf 140}, B516 (1965).
  %doi:10.1103/PhysRev.140.B516
  %%CITATION = doi:10.1103/PhysRev.140.B516;%%
  %612 citations counted in INSPIRE as of 03 Jun 2019

%\cite{He:2014laa}
\bibitem{He:2014laa} 
  T.~He, V.~Lysov, P.~Mitra and A.~Strominger,
  ``BMS supertranslations and Weinberg’s soft graviton theorem,''
  JHEP {\bf 1505}, 151 (2015)
  %doi:10.1007/JHEP05(2015)151
  [arXiv:1401.7026 [hep-th]].
  %%CITATION = doi:10.1007/JHEP05(2015)151;%%
  %278 citations counted in INSPIRE as of 03 Jun 2019
  
  %\cite{Strominger:2013jfa}
\bibitem{Strominger:2013jfa} 
  A.~Strominger,
  ``On BMS Invariance of Gravitational Scattering,''
  JHEP {\bf 1407}, 152 (2014)
  %doi:10.1007/JHEP07(2014)152
  [arXiv:1312.2229 [hep-th]].
  %%CITATION = doi:10.1007/JHEP07(2014)152;%%
  %331 citations counted in INSPIRE as of 03 Jun 2019
 
  
  %\cite{Strominger:2017zoo}
\bibitem{Strominger:2017zoo} 
  A.~Strominger,
  ``Lectures on the Infrared Structure of Gravity and Gauge Theory,''
  arXiv:1703.05448 [hep-th].
  %%CITATION = ARXIV:1703.05448;%%
  %200 citations counted in INSPIRE as of 03 Jun 2019


%\cite{Campiglia:2015yka}
\bibitem{Campiglia:2015yka} 
  M.~Campiglia and A.~Laddha,
  ``New symmetries for the Gravitational S-matrix,''
  JHEP {\bf 1504}, 076 (2015)
  %doi:10.1007/JHEP04(2015)076
  [arXiv:1502.02318 [hep-th]].
  %%CITATION = doi:10.1007/JHEP04(2015)076;%%
  %63 citations counted in INSPIRE as of 03 Jun 2019
  
  %\cite{Campiglia:2015qka}
\bibitem{Campiglia:2015qka} 
  M.~Campiglia and A.~Laddha,
  ``Asymptotic symmetries of QED and Weinberg’s soft photon theorem,''
  JHEP {\bf 1507}, 115 (2015)
  
  
  
 \bibitem{Bondi:1962px} 
  H.~Bondi, M.~G.~J.~van der Burg and A.~W.~K.~Metzner,
  ``Gravitational waves in general relativity. 7. Waves from axisymmetric isolated systems'',
  Proc.\ R.\ Soc.\ A {\bf 269}, 21 (1962).
  %doi:10.1098/rspa.1962.0161
  %%CITATION = doi:10.1098/rspa.1962.0161;%%
  %834 citations counted in INSPIRE as of 06 Aug 2017

\bibitem{Sachs:1962zza} 
  R.~Sachs,
  ``Asymptotic symmetries in gravitational theory'',
  Phys.\ Rev.\  {\bf 128}, 2851 (1962).
 % doi:10.1103/PhysRev.128.2851
  %%CITATION = doi:10.1103/PhysRev.128.2851;%%
  %267 citations counted in INSPIRE as of 02 Aug 2017

\bibitem{Sachs:1962wk} 
  R.~K.~Sachs,
  ``Gravitational waves in general relativity. 8. Waves in asymptotically flat space-times'',
  Proc.\ R.\ Soc.\ A {\bf 270}, 103 (1962).
 % doi:10.1098/rspa.1962.0206
  %%CITATION = doi:10.1098/rspa.1962.0206;%%
  
  
\bibitem{Newman:1966ub} 
  E.~T.~Newman and R.~Penrose,
  ``Note on the Bondi-Metzner-Sachs group'',
  J.\ Math.\ Phys.\  {\bf 7}, 863 (1966).
  
 %\cite{Barnich:2015jua}
\bibitem{Barnich:2015jua}
G.~Barnich, P.~H.~Lambert and P.~Mao,
``Three-dimensional asymptotically flat Einstein\textendash{}Maxwell theory,''
Class. Quant. Grav. \textbf{32}, no.24, 245001 (2015)
%doi:10.1088/0264-9381/32/24/245001
[arXiv:1503.00856 [gr-qc]].
%23 citations counted in INSPIRE as of 08 Oct 2020
  
\bibitem{Barnich:2013sxa} 
  G.~Barnich and P.~H.~Lambert,
  ``Einstein-Yang-Mills theory: Asymptotic symmetries'',
  Phys.\ Rev.\ D {\bf 88}, 103006 (2013).
  %doi:10.1103/PhysRevD.88.103006
  [arXiv:1310.2698 [hep-th]].
  
  
   \bibitem{Donnay:2015abr} 
  L.~Donnay, G.~Giribet, H.~A.~Gonzalez and M.~Pino,
  ``Supertranslations and Superrotations at the Black Hole Horizon'',
  Phys.\ Rev.\ Lett.\  {\bf 116}, 091101 (2016).
 % doi:10.1103/PhysRevLett.116.091101
  [arXiv:1511.08687 [hep-th]].
  
  
  \bibitem{Akhmedov:2017ftb} 
  E.~T.~Akhmedov and M.~Godazgar,
  ``Symmetries at the black hole horizon'',
  Phys.\ Rev.\ D {\bf 96}, 104025 (2017).
  %doi:10.1103/PhysRevD.96.104025
  [arXiv:1707.05517 [hep-th]]
  
   \bibitem{Cai:2016idg} 
  R.~G.~Cai, S.~M.~Ruan and Y.~L.~Zhang,
  ``Horizon supertranslation and degenerate black hole solutions'',
  J. High Eergy Physics. {\bf 09}, (2016) 163.
  %doi:10.1007/JHEP09(2016)163
  [arXiv:1609.01056 [gr-qc]].
  %%CITATION = doi:10.1007/JHEP09(2016)163;%%    
  
  %\cite{Eling:2016xlx}
\bibitem{Eling:2016xlx}
C.~Eling and Y.~Oz,
``On the Membrane Paradigm and Spontaneous Breaking of Horizon BMS Symmetries,''
JHEP \textbf{07}, 065 (2016)
%doi:10.1007/JHEP07(2016)065
[arXiv:1605.00183 [hep-th]].
%39 citations counted in INSPIRE as of 08 Oct 2020

  
 \bibitem{Maitra:2018saa} 
  M.~Maitra, D.~Maity and B.~R.~Majhi,
  ``Symmetries near a generic charged null surface and associated algebra: an off-shell analysis,''
  Phys.\ Rev.\ D {\bf 97}, no. 12, 124065 (2018)
  
  \bibitem{Maitra:2019eix}
  M.~Maitra, D.~Maity and B.~R.~Majhi,
  ``Near horizon symmetries, emergence of Goldstone modes and thermality,''
  Eur. Phys. J. Plus \textbf{135}, no.6, 483 (2020)
  %doi:10.1140/epjp/s13360-020-00451-3
  [arXiv:1906.04489 [hep-th]].
  
  \bibitem{Price:1986yy}
R.~H.~Price and K.~S.~Thorne,
``Membrane Viewpoint on Black Holes: Properties and Evolution of the Stretched Horizon,''
Phys. Rev. D \textbf{33}, 915-941 (1986)
%doi:10.1103/PhysRevD.33.915
%208 citations counted in INSPIRE as of 24 Aug 2020  

%\cite{Bredberg:2011jq}
\bibitem{Bredberg:2011jq}
I.~Bredberg, C.~Keeler, V.~Lysov and A.~Strominger,
``From Navier-Stokes To Einstein,''
JHEP \textbf{07}, 146 (2012)
%doi:10.1007/JHEP07(2012)146
[arXiv:1101.2451 [hep-th]].
%156 citations counted in INSPIRE as of 24 Aug 2020

%\cite{Parikh:1997ma}
\bibitem{Parikh:1997ma}
M.~Parikh and F.~Wilczek,
``An Action for black hole membranes,''
Phys. Rev. D \textbf{58}, 064011 (1998)
%doi:10.1103/PhysRevD.58.064011
[arXiv:gr-qc/9712077 [gr-qc]].
%153 citations counted in INSPIRE as of 24 Aug 2020 

%\cite{De:2018zxo}
\bibitem{De:2018zxo}
S.~De and B.~R.~Majhi,
``Fluid description of gravity on a timelike cut-off surface: beyond Navier-Stokes equation,''
JHEP \textbf{01}, 044 (2019)
%doi:10.1007/JHEP01(2019)044
[arXiv:1810.07017 [hep-th]].
%4 citations counted in INSPIRE as of 24 Aug 2020

\bibitem{De:2019wok}
S.~De, S.~Dey and B.~R.~Majhi,
``Effective metric in fluid-gravity duality through parallel transport: a proposal,''
Phys. Rev. D \textbf{99}, no.12, 124024 (2019)
%doi:10.1103/PhysRevD.99.124024
[arXiv:1901.05735 [hep-th]].

\bibitem{Dey:2020ogs}
S.~Dey, S.~De and B.~R.~Majhi,
``Gravity dual of Navier-Stokes equation in a rotating frame through parallel transport,''
Phys. Rev. D \textbf{102}, no.6, 064003 (2020)
%doi:10.1103/PhysRevD.102.064003
[arXiv:2002.06801 [hep-th]].

%\cite{Penna:2015gza}
\bibitem{Penna:2015gza}
R.~F.~Penna,
``BMS invariance and the membrane paradigm,''
JHEP \textbf{03}, 023 (2016)
%doi:10.1007/JHEP03(2016)023
[arXiv:1508.06577 [hep-th]].
%34 citations counted in INSPIRE as of 25 Sep 2020

\bibitem{Booth:2012xm} 
  I.~Booth,
  ``Spacetime near isolated and dynamical trapping horizons,''
  Phys.\ Rev.\ D {\bf 87}, no. 2, 024008 (2013)
  %doi:10.1103/PhysRevD.87.024008
  [arXiv:1207.6955 [gr-qc]].
  %%CITATION = doi:10.1103/PhysRevD.87.024008;%%
  %29 citations counted in INSPIRE as of 17 Oct 2019
  
  \bibitem{MORALES}
  E.~M.~Morales,
  ``On a Second Law of Black Hole Mechanics in a Higher Derivative Theory of Gravity,''
  http://www.theorie.physik.uni-goettingen.de/forschung/qft/theses/dipl/Morfa-Morales.pdf.
  
  %\bibitem{Barnich:2010eb} 
  %\cite{Barnich:2010eb}
\bibitem{Barnich:2010eb}
G.~Barnich and C.~Troessaert,
``Aspects of the BMS/CFT correspondence,''
JHEP \textbf{05}, 062 (2010)
%doi:10.1007/JHEP05(2010)062
[arXiv:1001.1541 [hep-th]].
%327 citations counted in INSPIRE as of 08 Oct 2020
  
  
    
\bibitem{Paddy}
  T.~Padmanabhan,
  ``Gravitation: Foundations and Frontiers'', (Cambridge University Press. Cambridge, England, 2010) Chap 8, p.394.
  
  
   %\cite{Tolman:1930zza}
  \bibitem{Tolman:1930zza}
  R.~C.~Tolman,
``On the Weight of Heat and Thermal Equilibrium in General Relativity,''
  Phys. Rev. \textbf{35}, 904-924 (1930)
  %doi:10.1103/PhysRev.35.904
   
   %\cite{Tolman:1930ona}
  \bibitem{Tolman:1930ona}
  R.~Tolman and P.~Ehrenfest,
 ''Temperature Equilibrium in a Static Gravitational Field,''
  Phys. Rev. \textbf{36}, no.12, 1791-1798 (1930)
  %doi:10.1103/PhysRev.36.1791
  
\bibitem{Tolman}  
  R.~ C.~Tolman,
   “Thermodynamics and Relativity. I”,
   Science \textbf{77}, 1995, 291-298 (1933).
  

   R. C. Tolman, “Thermodynamics and Relativity. II”,
   Science \textbf{77}, 1996, 313-317 (1933).
   
   %\cite{Santiago:2018kds}
  \bibitem{Santiago:2018kds}
 J.~Santiago and M.~Visser,
``Tolman temperature gradients in a gravitational field,''
 Eur. J. Phys. \textbf{40}, no.2, 025604 (2019)
 %doi:10.1088/1361-6404/aaff1c
[arXiv:1803.04106 [gr-qc]]

\bibitem{Komar:1958wp}
A.~Komar,
``Covariant conservation laws in general relativity,''
Phys. Rev. \textbf{113}, 934-936 (1959)
%doi:10.1103/PhysRev.113.934

\bibitem{Banerjee:2010yd}
R.~Banerjee and B.~R.~Majhi,
``Statistical Origin of Gravity,''
Phys. Rev. D \textbf{81}, 124006 (2010)
%doi:10.1103/PhysRevD.81.124006
[arXiv:1003.2312 [gr-qc]].

\bibitem{Banerjee:2010ye}
R.~Banerjee, B.~R.~Majhi, S.~K.~Modak and S.~Samanta,
``Killing Symmetries and Smarr Formula for Black Holes in Arbitrary Dimensions,''
Phys. Rev. D \textbf{82}, 124002 (2010)
%doi:10.1103/PhysRevD.82.124002
[arXiv:1007.5204 [gr-qc]].
   
%\cite{Santiago:2018lcy}
\bibitem{Santiago:2018lcy}
J.~Santiago and M.~Visser,
``Tolman-like temperature gradients in stationary spacetimes,''
Phys. Rev. D \textbf{98}, no.6, 064001 (2018)
%doi:10.1103/PhysRevD.98.064001.
[arXiv:1807.02915 [gr-qc]].

% \bibitem{Buchdahl}  
%  A.~ Buchdahl,
% “Temperature equilibrium in a stationary gravitational field”, 
 % Phys. Rev. 76 (1949) 427
 
 \end{thebibliography}
\end{document}